\documentclass[a4paper,fleqn,usenatbib]{mnras}

\usepackage{times}
\usepackage{amsmath,amssymb}
\usepackage{graphicx}
\usepackage{hyperref}
\usepackage{breakurl}

\title[Early life of massive galaxy progenitors]{
Young and turbulent: the early life of massive galaxy progenitors
}

\author[Fiacconi et al.]{Davide Fiacconi$^{1,2,}$\thanks{E-mail: fiacconi@ast.cam.ac.uk}, 
Lucio Mayer$^{2}$,
Piero Madau$^{2,3,4}$,
Alessandro Lupi$^{4}$,
\newauthor
Massimo Dotti$^{5,6}$,
\& Francesco Haardt$^{6,7}$
\\
$^{1}$Institute of Astronomy and Kavli Institute for Cosmology, University of Cambridge, Madingley Road, Cambridge CB3 0HA, UK\\
$^{2}$Center for Theoretical Astrophysics and Cosmology, Institute for Computational Science, University of Zurich,\\
Winterthurerstrasse 190, CH-8057 Z\"{u}rich, Switzerland\\
$^{3}$Department of Astronomy \& Astrophysics, University of California Santa Cruz, 1156 High St., Santa Cruz, CA 95064, USA\\
$^{4}$Institut d'Astrophysique de Paris, Sorbonne Universit\'es, UPMC Univ Paris 6 et CNRS, UMR 7095, 98 bis bd Arago, 75014 Paris, France\\
$^{5}$Dipartimento di Fisica, Universit\`a di Milano-Bicocca, Piazza della Scienza 3, I-20126 Milano, Italy\\
$^{6}$INFN, Milano-Bicocca, P.za della Scienza 3, I-20126 Milano, Italy\\
$^{7}$DiSAT, Universit\`a degli Studi dell'Insubria, Via Valleggio 11, I-22100 Como, Italy
}

\begin{document}

\date{\today}

\pagerange{\pageref{firstpage}--\pageref{lastpage}} \pubyear{2016}

\maketitle

\label{firstpage}


\begin{abstract}
We present results from the ``Ponos'' simulation suite on the early evolution of a massive, 
$M_{\rm vir}(z=0)=1.2\times 10^{13}$~M$_{\sun}$ galaxy.
At $z\gtrsim6$, before feedback from a central supermassive black hole becomes dominant, the main galaxy has a stellar mass $\sim 2\times 10^{9}$~M$_{\sun}$ and a star formation rate $\sim 20$~M$_{\sun}$~yr$^{-1}$.
The galaxy sits near the expected main sequence of star-forming galaxies at those redshifts, and resembles moderately star-forming systems observed at $z>5$.
The high specific star formation rate results in vigorous heating and stirring of the gas by supernovae feedback, and the galaxy develops a thick and turbulent disc, with gas velocity dispersion $\sim 40$~km~s$^{-1}$, rotation to dispersion ratio $\sim 2$, and with a significant amount of gas at $\sim 10^5$~K.
The Toomre parameter always exceeds the critical value for gravito-turbulence, $Q\sim 1.5-2$, mainly due to the contribution of warm/hot gas inside the disc.
Without feedback, a nearly gravito-turbulent regime establishes with similar gas velocity dispersion and lower $Q$. 
We propose that the ``hot and turbulent'' disc regime seen in our simulations, unlike the ``cold and turbulent'' gravito-turbulent regime of massive clumpy disc galaxies at $z\sim 1-2$, is a fundamental characterisation of main sequence galaxies at $z\gtrsim 6$, as they can sustain star formation rates comparable to those of low-mass starbursts at $z=0$.
This results in no sustained coherent gas inflows through the disc, and in fluctuating and anisotropic mass transport, possibly postponing the assembly of the bulge and causing the initial feeding of the central black hole to be highly intermittent.
\end{abstract}



\section{Introduction}

Since the early 90's, the growing number of galaxy surveys exploring different wavelengths and cosmological volumes have provided an immense body of data to reconstruct the formation and evolution of galaxies (e.g. \citealt{york+00,steidel+03,lefevre+05,lilly+07,walter+08,baldry+10,grogin+11,koekemoer+11,kochanek+12}).
At the same time, the $\Lambda$CDM cosmological paradigm of structure formation has produced detailed predictions on the statistical matter distribution in the Universe
(e.g. \citealt{percival+01,springel+05,reed+07,viel+09,percival+10,reid+10,klypin+11}),
but the theory of galaxy formation within this framework is still developing, through
analytical work (e.g. \citealt{white+78,blumenthal+84,mo+98}) and semi-analytic models (e.g. \citealt{kauffmann+93, cole+94,somerville+99,springel+05,bower+06,croton+06,guo+11,henriques+13}), as well as cosmological numerical simulations (e.g. \citealt{diemand+08,crain+09,schaye+10,agertz+11,guedes+11,dubois+14,hopkins+14,vogelsberger+14,schaye+15}; and references therein).

Most of the theoretical successes have been achieved at low to moderate redshifts ($z\sim 0 - 1.5$), where the majority and the highest quality data are available.
The latter have been fundamental to gauge the characteristics and the parameters of phenomenological, sub-resolution recipes introduced in numerical simulations (as well as in semi-analytic models) in order to model physical processes on scales that are currently inaccessible directly, such as gas radiative cooling, gas chemistry evolution, star formation, stellar feedback, black hole accretion and feedback (e.g. \citealt{stinson+06,dallavecchia+08,gnedin+09,wiersma+09,vogelsberger+13,keller+14}).
Large-box cosmological simulations, despite suffering from relatively low resolution, have been able to statistically reproduce the main properties of different galaxy populations, from small, star-forming irregular galaxies to massive quiescent ellipticals \citep{vogelsberger+14,schaye+15}.
On the other hand, high-resolution zoom-in simulations focussing on the formation and growth of a limited number of systems have also satisfactory modelled the formation and growth of dwarf galaxies \citep[e.g.][]{governato+10}, Milky-Way-like hosts \citep[e.g.][]{agertz+11,guedes+11}, as well as massive ellipticals at the centre of galaxy groups \citep[e.g.][]{feldmann+10,feldmann+15,fiacconi+15}.

At higher redshifts $z \sim 2$, recent observational campaigns have boosted the interest in the population of massive star forming galaxies that show peculiar properties when compared to local counterparts of similar size.
Such systems are often characterised by massive discs with baryonic mass $\sim 10^{11}$~M$_{\sun}$ and star formation rates as high as $\sim 100$~M$_{\odot}$~yr$^{-1}$, with a turbulent interstellar medium that accounts for $\gtrsim 30 \%$ of the baryonic mass (e.g. \citealt{genzel+06,foster-schreiber+09,daddi+10,tacconi+10,wisnioski+15}).
The new observations have triggered lively discussions in the theoretical community, also requiring the development of new theoretical ideas to explain the observations and to make new predictions (e.g. violent disc instability; \citealt{mandelker+14,inoue+16}).

Pushing it forward, the very high redshifts ($z \gtrsim 4$) still remain a partially unexplored territory, clearly because of the larger technical difficulties involved in new, cutting-edge observations.
Nonetheless, galaxies at $z > 4$ have been detected in a few different ways.
Optical and near infra-red observations have targeted star-forming galaxies by identifying them through the flux dropout in adjacent bands around the Lyman break (e.g. \citealt{madau+96,steidel+99,bouwens+03,oesch+10}).
Those have permitted to constrain the evolution of the cosmic star formation rate density as well as of the ultraviolet (UV) luminosity function of star forming galaxies out to $z \gtrsim 8-10$, showing that the latter has a steep low luminosity tail, when compared to local samples (e.g. \citealt{bouwens+07,bouwens+11,oesch+12,oesch+14,bouwens+15}).
Moreover, other wavelengths have been effective in providing information about the early galaxy population.
Sub-millimetre galaxies are an example of massive (stellar mass $\sim 10^{11}$~M$_{\sun}$), highly star-forming (star formation rates $\gtrsim 500$~M$_{\sun}$~yr$^{-1}$), dusty galaxies that have been detected mostly at $z > 2.5-3$ (e.g. \citealt{chapman+05,younger+09,casey+14}).

While those studies have been important to understand the global properties of the first galaxies in the Universe, they have mostly revealed the luminous tail of the galaxy population and they are still not able to characterise their structure (but see \citealt{oesch+10b,debreuck+14}).
Nonetheless, both available facilities, such as the Hubble Space Telescope (HST) or the Very Large Telescope (VLT), as well as new observatories that recently came online, such as the Atacama Large Millimeter Array (ALMA), are starting to discover smaller and possibly more typical galaxies at $z \geq 6$.
For example, \citet{bradley+12} have used HST imaging to identify a few Lyman break galaxy candidates at $z \approx 7$, lensed by the foreground galaxy cluster A1703.
The most luminous likely has a stellar mass $\sim 10^{9}$~M$_{\sun}$ and a star formation rate $\sim 8$~M$_{\sun}$~yr$^{-1}$.
Similarly, \citet{watson+15} have combined HST, VLT, and ALMA observations to constrain the stellar mass, star formation and gas content of another highly magnified Lyman break galaxy beyond the Bullet cluster.
They also find a stellar mass $\sim 2 \times 10^{9}$~M$_{\sun}$, a star formation rate $\sim 10$~M$_{\sun}$~yr$^{-1}$, and a gas fraction $\sim 40-50\%$, all confined within a physical surface $\sim 1.5$~kpc$^2$.
All these objects typically have specific star formation rates $\sim5$-10~Gyr$^{-1}$ (see also e.g. \citealt{tasca+15}).

These recent observations, combined with those from the next generation of both space-based (James Webb Space Telescope; JWST) and earth-based (e.g. European Extremely Large Telescope; E-ELT) telescopes, require new interpretations and predictions on the theoretical side, where less has been done compared to low/medium redshifts, except regarding the most massive population of galaxies at $z > 6-8$ (e.g. \citealt{choi+12,oshea+15,paardekooper+15,ocvirk+16,waters+16}).
Motivated by this, we investigate the early phases of the formation and the evolution of a galaxy that becomes a massive elliptical at $z=0$.
We focus on the first burst of star formation, prior to the assembly of the central supermassive black hole and before active galactic nuclei (AGN) feedback becomes dominant.
We use a new high-resolution hydrodynamic run that is part of the recent Ponos program of zoom-in cosmological simulations of massive galaxies \citep{fiacconi+16}.
Our new simulation reproduces the main features of recently observed star forming galaxies at $z \sim 7$, and allows us to study in details the properties of the early interstellar medium that drive the star formation and may determine the early feeding habits of the central black hole.

The paper is organised as follows.
In Section \ref{sec_2}, we describe our numerical techniques and the features of the PonosHydro simulation.
We describe our main results in Section \ref{sec_3}, focussing on the properties of the interstellar medium and the early star formation history of the simulated system at $z\sim 6$.
We present our conclusions in Section \ref{sec_4}, cautioning the reader about the limitations of our results and discussing the possible implications of our findings.
In the following, when not explicitly specified, all lengths and densities are given in physical units.


\section{Methods} \label{sec_2}


\subsection{Initial conditions}

We perform and analyse a new simulation that complements the suite of cosmological, zoom-in simulations presented by \citet{fiacconi+16} and named ``Ponos''.
This is a high-resolution version of the run focusing on the halo originally dubbed as ``PonosV''.
This new version includes hydrodynamics and self-consistent baryonic physics, and we refer to it as ``PH'' or ``PonosHydro'' in the following.
The target halo evolves in a box of of 85.5 comoving Mpc and it reaches a mass $\approx 1.2 \times 10^{13}$~M$_{\sun}$ at $z = 0$.
We adopt a $\Lambda$CDM cosmology consistent with the results of \emph{Wilkinson Microwave Anisotropy Probe} 7/9 years, parametrised by $\Omega_{\rm m,0}=0.272$, $\Omega_{\Lambda,0}=0.728$, $\Omega_{\rm b,0}=0.0455$, $\sigma_{8} = 0.807$, $n_{\rm s} = 0.961$, and $H_0 = 70.2$~km~s$^{-1}$~Mpc$^{-1}$ \citep{komatsu+11,hinshaw+13}.
The original initial conditions (ICs) are part of the AGORA\footnote{\url{https://sites.google.com/site/santacruzcomparisonproject/home}} code-comparison project \citep{kim+14}.
We generate new ICs of the same halo using the {\sc Music}\footnote{\url{http://www.phys.ethz.ch/~hahn/MUSIC/}} code \citep{hahn+11}.
Our ICs are optimised to follow the growth of the halo until $z=6$.
They consist of a base cube of 128$^3$ particles per side starting at $z=100$, with additional nested levels of refinement to increase the resolution within the Lagrangian region that maps the particles contained within $2.5 R_{\rm vir}$ at $z=6$ on to the ICs.
We define the virial radius $R_{\rm vir}$ as the spherical radius that encompasses a mean matter density $\Delta(z) \rho_{\rm c}(z)$, where $\rho_{\rm c}(z)$ is the critical density that the Universe needs to be flat, and $\Delta(z)$ is the $z$-dependent virial over-density defined by \citet{bryan+98}.
As a consequence, the virial mass is defined as $M_{\rm vir} = 4 \pi \Delta(z) \rho_{\rm c}(z) R_{\rm vir}^3 / 3$.
At $z=6$, the virial mass of the main galaxy is $\approx 1.5 \times 10^{11}$~M$_{\sun}$.
In fact, we use the following procedure to determine the high-resolution region on the ICs \citep{fiacconi+16}: (i) we run a $128^3$, dark-matter-only, full-box simulation to identify the main halo at $z=6$; (ii) we trace the particles within 2.5$R_{\rm vir}$ back on to the ICs; (iii) we locally increase the resolution of the ICs by adding one level of refinement within a rectangular box that contains all the identified particles; (iv) we evolve again the ICs until $z=6$; (v) we repeat (ii) and we add two additional levels of refinement within the convex hull that contains all the identified particles.
Every level of refinement increases the spatial and mass resolution by a factor 2 and 8, respectively.
We iterate steps (iv) and (v) until we add 7 additional levels of refinement above the base cube, introducing gas particles in the last level.
We verified with dark-matter-only test runs and in the main run that this procedure allows us to maintain the fraction of contaminating particles\footnote{Here the term ``contaminating particles'' refers to dark matter particles from coarser levels of refinement in the ICs.} in both mass and number well below 0.1\% within the virial volume at all $z\geq 6$.
Finally, the highest resolution dark matter and gas particle have a mass $m_{\rm dm} = 4397.6$~M$_{\sun}$ and $m_{\rm g} = 883.4$~M$_{\sun}$, respectively.
The force resolution is determined by the softening length, which is set to 1/60 of the mean particle separation at each level of refinement.
This corresponds to a minimum dark matter softening $\epsilon_{\rm dm} = 81.8$ physical pc and to a gas softening $\epsilon_{\rm g} = 47.9$ physical pc.
The softening is kept constant in physical units during evolution at $z<9$, while it remains constant in comoving coordinates at higher redshifts.
The total number of particles in the ICs is 118,694,002, while the total number of particles within the viral radius at $z=6.5$ is 56,213,155.


\subsection{Simulation code}\label{sec_sim_code}

We evolve the simulation using the {\sc gasoline} code \citep{wadsely+04}.
The code computes gravitational interactions using a 4$^{\rm th}$-order (i.e. hexadecapole) multipole expansion of the gravitational force on a KD binary tree, following the original scheme adopted by the {\sc pkdgrav} code \citep{stadel+01,stadel+13}.
{\sc gasoline} models the gas dynamics using the smoothed particle hydrodynamics (SPH) algorithm (e.g. \citealt{lucy+77,gingold+77,hernquist+89}).
In addition to the standard SPH formulation, the energy equation includes a term for thermal energy and metal diffusion (with a coefficient $C = 0.05$) as introduced and discussed by \citet{wadsely+08} and \citet{shen+10}.
This approach reduces the artificial surface tension that arises close to strong density gradients where Kelvin-Helmholtz instability  may develop (e.g. \citealt{agertz+07,read+12,hopkins+13,keller+14}).

{\sc gasoline} includes several sub-resolution models to treat the radiative cooling of the gas, the formation of stars and the impact of their feedback in terms of supernovae-injected energy and mass released by winds.
The gas is allowed to cool radiatively in the optically-thin limit by solving the non-equilibrium reaction network of HI, HII, HeI, HeII and HeIII.
We take into account the contribution from metal lines with a temperature floor $T_{\rm floor} \approx 100$~K following \citet{shen+13}.
The simulation also includes a uniform, redshift-dependent ultraviolet radiation background due to stellar and quasar reionisation according to \citet{haardt+12}.
The implementation of star formation and stellar feedback mostly follows the prescriptions from \citet{stinson+06}.
Stars form when: (i) the local density exceeds the threshold ${\rm n}_{\rm SF} = 10$~H~cm$^{-3}$ (ii) the local temperature goes below $T_{\rm SF} = 10^4$~K; (Iii) the local gas over-density is $>2.64$; and (iv) the flow is convergent and locally Jeans unstable.
When the above criteria are fulfilled, stars form according to a Schmidt-like low, $\dot{\rho}_{\star} = \epsilon_{\rm SF} \rho_{\rm g} / t_{\rm dyn} \propto \rho_{\rm g}^{3/2}$, where $\rho_{\star}$ is the mass density of formed stars, $\rho_{\rm g}$ is the local gas mass density, and $t_{\rm dyn} = 1/\sqrt{G \rho_{\rm g}}$ is the local dynamical time.
The efficiency $\epsilon_{\rm SF} = 0.05$ is a phenomenological parameter meant to capture the average star-formation efficiency.
Each stellar particle has an initial mass $m_{\star} = 0.4 m_{\rm g} = 353.4$~M$_{\sun}$ and represents a stellar population with a \citet{kroupa+01} initial mass function.

We choose ${\rm n}_{\rm SF}$ to be about the density reached when resolving the local Jeans mass with at least one kernel at the lowest temperature reached by the simulation, i.e. the temperature floor of the cooling function, namely:
\begin{equation}
{\rm n}_{\rm SF} \lesssim 16.4~\left( \frac{T_{\rm floor}}{100~{\rm K}} \right)^{3} \left( \frac{N}{32} \right)^{-2} \left( \frac{m_{\rm g}}{883.4~{\rm M}_{\sun}} \right)^{-2}~{\rm H~cm^{-3}},
\end{equation}
where $N$ is the number of gas particles per kernel.
This choice ensures that, even in the most extreme case (i.e. gas at the lowest temperature that has not formed stars yet), the local collapse is resolved.
However, since $T_{\rm SF} \gg T_{\rm floor}$, the Jeans mass during collapses that lead to star-formation episodes is likely resolved with a number of particles $\gg N$.
Indeed, at ${\rm n}_{\rm SF}$ and $T_{\rm SF}$, the local Jeans mass is resolved with $\sim 1220$ kernel masses, which corresponds to $\sim 39,000$ particles in the regions at the verge of the collapse that might form stars.

\begin{table}
\caption{List of performed simulations and of their main features.
From left to right: label of the simulation, initial redshift of the simulation, final redshift of the simulation, $\alpha$ parameter of the pressure floor (see the text), initial mass of gas particles, mass of the lightest dark matter particles, gravitational softening of the gas, smallest gravitational softening of the dark matter.}
\label{tab:summmary}
\begin{tabular}{lccccccc}
\hline
Label & $z_{\rm ini}$ & $z_{\rm end}$ & $\alpha$ & $m_{\rm g}$ & $m_{\rm dm}$ & $\epsilon_{\rm g}$ & $\epsilon_{\rm dm}$ \\
 &  &  &  & (M$_{\sun}$) & (M$_{\sun}$) & (pc) & (pc) \\
\hline
PH & 100 & 6.5 & 9 & 883.4 & 4397.6 & 47.9 & 81.8 \\
PH\_PF1 & 8 & 6.5 & 1 & 883.4 & 4397.6 & 47.9 & 81.8 \\
PH\_NF$^{a}$ & 8 & 7.1 & 9 & 883.4 & 4397.6 & 47.9 & 81.8 \\
PH\_LR & 100 & 6.5 & 9 & 7067.2 & 35180.8 & 95.8 & 163.6 \\
\hline
\end{tabular}

$^{a}$ This run does not include neither star formation nor stellar feedback.
\end{table}

\begin{figure*}
\begin{center}
\includegraphics[width=2\columnwidth]{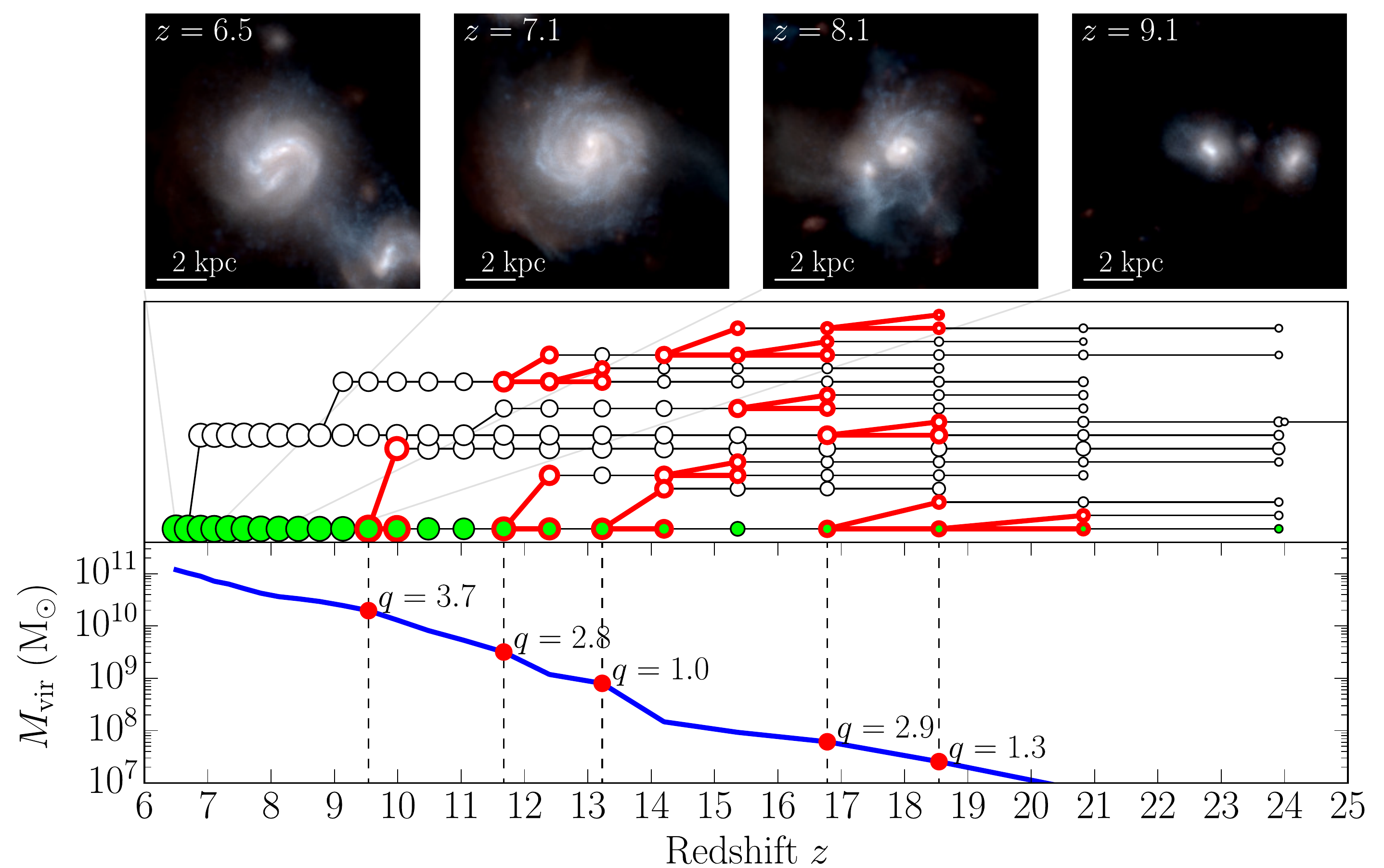}
\caption{
Upper row, from left to right: sequence of images at $z=6.5$, 7.1, 8.1 and 9.1, respectively, of the stellar component of the main galaxy in the restframe U, V and J bands.
Middle row: merger tree of the main halo as a function of redshift.
The size of each circle is proportional to the logarithm of the virial mass.
Green circles denote the main branch of the merger tree, while thick, red circles connected by thick, red lines mark the occurrence of a major merger with mass ratio $q < 4$.
Lower panel: evolution of the virial mass of the main halo as a function of redshift.
Red dots and vertical dashed lines show major mergers together with their mass ratio $q$.
The main halo reaches about $10^{11}$~M$_{\sun}$ by $z \gtrsim 6$.}
\label{fig_merger_tree}
\end{center}
\end{figure*}

During their lifetime, light stars with mass between 1 and 8~M$_{\sun}$ release $\sim 40 \%$ of the particle mass and metals to the surrounding gas through winds following \citet{weidemann+87}.
On the other hand, massive stars between 8 and 40~M$_{\sun}$ are responsible for Type II supernovae, which inject energy, mass and metals to the surrounding gas following the analytical blast wave solution of \citet{mckee+77}.
In particular, each Type II supernova releases $10^{51}$~erg in thermal energy to the gas particles within the maximum radius that the blast wave can reach.
The cooling of those gas particles is temporarily turned off for the time corresponding to the snowplough phase of the blast wave.
Type Ia supernovae also contribute to the supernova feedback budget injecting the same energy as Type II supernovae, but without the shut off of the cooling. 
They releases a mass of 1.4~M$_{\sun}$, where 0.63 and 0.13~M$_{\sun}$ are constituted by iron and oxygen, respectively.
Their frequency is given by the binary fraction estimated by \citet{raiteri+96}.

We also adopt the pressure floor described by e.g. \citet{roskar+15} in order to avoid spurious fragmentation.
Specifically, the minimum pressure of each gas particle (at a given density and temperature) is $P_{\rm min} = \alpha \max(\epsilon_{\rm g}, h)^2 G \rho^2$, where $h$ is the local SPH smoothing length, $\rho$ is the gas mass density, and $\alpha = 9$ is a safety factor.
This choice of $\alpha$ is such that the local Jeans length is always resolved with $N_{\lambda} \approx 6$ ($N_{\lambda} \approx 3$) resolution elements $\max(\epsilon_{\rm g}, h)$, assuming that the Jeans length is defined as $\lambda_{\rm J} = c_{\rm s} \sqrt{\pi / (G \rho)}$ ($\lambda_{\rm J} = \sqrt{15 k_{\rm B} T / (4 \pi G \rho \mu m_{\rm p})}$).
The local sound speed of the gas is $c_{\rm s}= \sqrt{\gamma k_{\rm B} T / (\mu m_{\rm p})}$, where $\gamma = 5/3$ is the adiabatic index, $k_{\rm B}$ is the Boltzmann constant, $\mu \approx 0.6$ is the mean molecular weight (assuming that most of the gas is ionised), and $m_{\rm p}$ is the proton mass.
The criterion on the Jeans length translates in the local Jeans mass resolved with $\approx (\epsilon_{\rm g} / h)^{3} N_{\lambda}^{3} / 8 \approx 15.6~N_{\lambda}^{3}$ kernel masses, where we substitute $\epsilon_{\rm g} / h = 5$ in the last equality after noting that the distribution of the ratio $h/\epsilon_{\rm g}$ within the main galaxy is peaked at about 0.2.
We note that we adopt a pressure floor to avoid numerical fragmentation, while we leave the temperature of the gas free to evolve  according to the energy equation and the effect of radiative cooling and heating (e.g. \citealt{richings+15}).
This means that the equation of state of the gas does not effectively follow an ideal gas law anymore when relating the pressure to the density and the temperature.
The usage of the pressure floor also imply that the gravitational collapse is at least partially suppressed for structures initially collapsing on scales below $N_{\lambda} \epsilon_{\rm g} \sim 150-300$~pc \citep{bate+97}.
However, we have checked that the gas phase-space region where the chosen pressure floor kicks in overlaps with the conditions for forming stars only at high densities and low temperatures. 
This implies that the initial phases of gravitational collapses that lead to star formation are physical and well resolved, while the pressure floor ensures that spurious fragmentation on smaller scales is avoided.
Nonetheless, this may have a dynamical role in the evolution of the interstellar medium of the simulated galaxy; therefore, we also run an additional run from $z \approx 8$ (after that the disc has re-formed; see Section \ref{sec_3}) to $z=6.5$ adopting a lower pressure floor with $\alpha = 1$.
This additional run is dubbed ``PH\_PF1''.
We also restart an additional version of run PH from $z \approx 8$ to $z = 7.1$, including radiative cooling but without star formation and stellar feedback, in order to test the impact of feedback on the properties of the interstellar medium (see Section \ref{sec_ism}). We refer to this run as ``PH\_NF''.

In addition to run PH, we have also simulated a lower resolution version, i.e.  ``PH\_LR'', adopting the same parameters.
This run has a mass and force resolution 8 times and 2 times coarser than the main run, respectively.
The initial total number of particle is 17,952,072, while the main halo contains about 5,974,677 particles within the virial radius at $z=6.5$.
We use this run for resolution tests that we show in Appendix \ref{appendix_resolution_tests}.
We summarise the labelling and the main features of all the simulations in Table \ref{tab:summmary}.


\subsection{Halo detection} \label{halo_detection}

We identify dark matter haloes (and then the contained galaxies) using the {\sc amiga halo finder} \citep{gill+04, knollmann+09}.
Every halo is defined as a gravitationally bound group of at least 100 particles within a virial radius $R_{\rm vir}$.
Then we construct the merger tree of the main halo matching the dark matter particle IDs among every two snapshots from $z=6.5$ to $z \simeq 30$ (e.g. \citealt{fiacconi+15}).
The main progenitor branch is determined as the progenitor halo that maximises the number $f_{\rm shared} = N_{\rm shared} / \sqrt{N_{\rm h} N_{\rm prog}}$ through each snapshot, where $N_{\rm shared}$ is number of particles shared among the halo and its progenitor, while $N_{\rm h}$ and $N_{\rm prog}$ are the particles of the halo and the progenitor, respectively \citep{fiacconi+15}.


\section{Results} \label{sec_3}


\begin{figure}
\begin{center}
\includegraphics[width=\columnwidth]{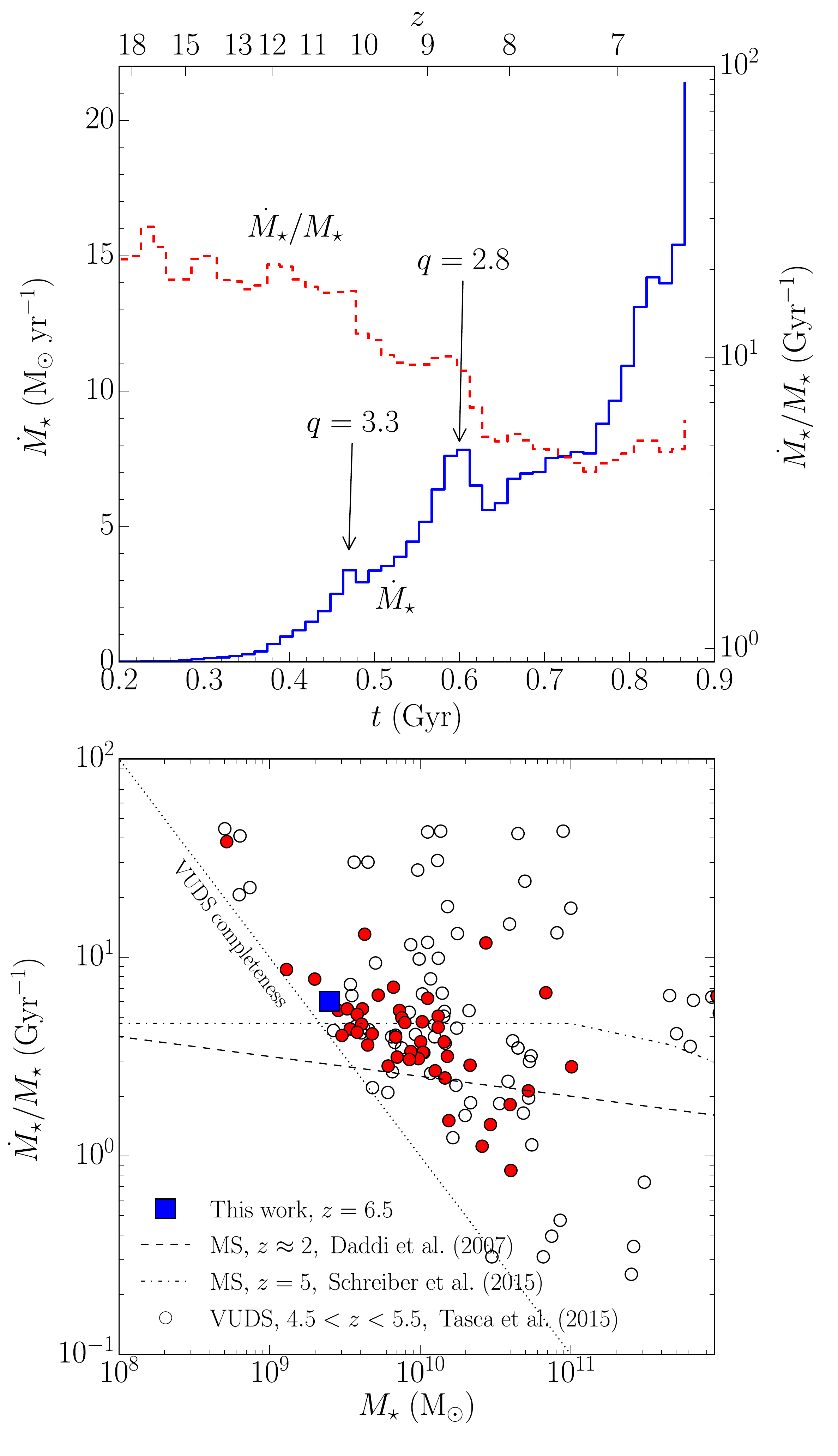}
\caption{
Upper panel: time evolution of the star formation rate ($\dot{M}_{\star}$, blue continuous line, left $y$-axes) and of the specific star formation rate ($\dot{M}_{\star} / M_{\star}$, red dotted line, right $y$-axes).
The latest two major mergers of the main halo are highlighted with their mass ratio $q$.
Lower panel: specific star formation rate as a function of the stellar mass for our simulation at $z=6.5$ (PH, blue square) and for the data from the VUDS survey in the redshift range $4.5 < z < 5.5$  \citep{tasca+15}.
The VUDS data are represented as red filled and empty circles when the redshift determination is $\sim 100 \%$ and $\sim 70-75\%$ reliable, respectively.
The thin dotted line shows the VUDS completeness limit in the determination of the star formation rates.
The dashed and dot-dashed lines show the determination of the main sequence of star forming galaxies at $z \approx 2$ from \citet{daddi+07} and the analytical fit of the main sequence by \citet{schreiber+15} extrapolated to $z=5$, respectively.}
\label{fig_SFH}
\end{center}
\end{figure}

\subsection{Evolution of the main galaxy}

The global evolution of the main halo of the run PH is summarised in Figure \ref{fig_merger_tree}.
Specifically, we show the merger tree of the main halo, highlighting the major mergers with mass ratio $q < 4$ (where we define $q \equiv M_{1} / M_{2} > 1$), and the growth of the virial mass $M_{\rm vir}$ with redshift.
The early growth of the main halo is characterised by a sequence of several major mergers with mass ratios between $q=1$ and $q=3$ from $z\approx 18.5$ to $z\approx 11.6$.
During this phase, $M_{\rm vir}$ quickly grows from $< 10^8$~M$_{\sun}$ to $\sim 4 \times 10^{9}$~M$_{\sun}$ through the rapid accretion of both dark matter and gas, also triggered by the major mergers.
By $z\approx 11$, the main halo has formed, i.e. it has reached about 5\% of its final mass at $z=6.5$, which is $M_{\rm vir} \simeq 1.2 \times 10^{11}$~M$_{\sun}$.
Slightly later, at $z\approx 10$, the virial radius of the main halo ($R_{\rm vir} \approx 7.7$~kpc in physical units) begins to overlap with the virial volume of a nearby halo with mass 3.7 times smaller than the main one.
This is the last major merger of the main halo during the time span that we have simulated.
It completes by $z \sim 9$, though the two central galaxies merge slightly after at $z\approx 8$, as shown by Figure \ref{fig_merger_tree}.
Such a time lag between the merger of the two haloes and their central galaxies, $\sim 100$~Myr from $z=9$ to $z=8$, is consistent with the dynamical friction timescale for the central galaxies to sink at the centre of the remnant halo after a $q=4$ merger, namely $\sim 0.2~t_{\rm H}(z) \sim 170$~Myr, where $t_{\rm H}(z) = H^{-1}(z)$ is the Hubble time at $z$ and the last equality is obtained assuming $z=9$ \citep{krivitsky+97,boylan-kolchin+08,jiang+08,hopkins+10}.

\begin{figure*}
\begin{center}
\includegraphics[width=2\columnwidth]{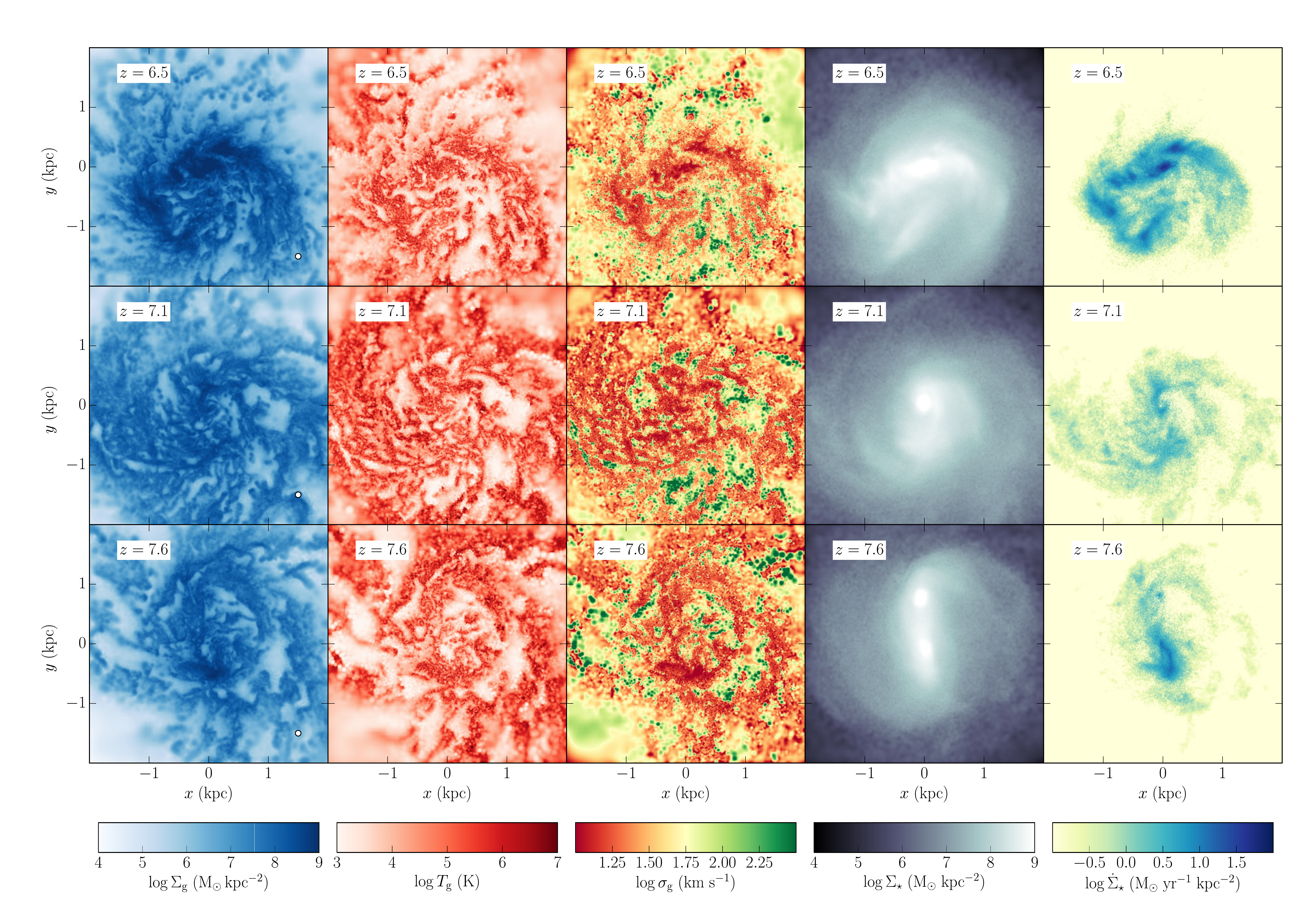}
\caption{
Evolution of the galactic disc in run PH between $z=7.6$ and $z=6.5$ (from bottom to top).
From left to right: gas surface density, local gas temperature in the disc mid-plane, local gas velocity dispersion (computed locally as an SPH weighted average) in the disc mid-plane, stellar surface density, and star formation rate surface density.
All quantities are measured within a 200~pc thick slice centred on the disc plane.
The star formation rate surface density is determined from the surface density of stellar particles younger than 10~Myr.
The circles in the lower-right corners of the first column have radii equal to the gravitational softening of the gas, 47.9 physical pc.
}
\label{fig_disc_faceon}
\end{center}
\end{figure*}

After the last major merger, the galaxy quickly builds an extended stellar and gaseous disc at $z\approx 7.5$.
This is visible e.g. in the mock UVJ image at $z=7.1$ in Figure \ref{fig_merger_tree}.
Those images are obtained similarly to \citet{fiacconi+15}: we consider each stellar particle as a stellar population with a \citet{kroupa+01} initial mass function of age $\tau$ and metallicity $Z$ and we determine its total luminosity by interpolating a table based on the synthetic stellar population models of the Padova group \citep{marigo+08,girardi+10,bressan+12}.
These tables\footnote{Similar tables can be obtained at \url{http://stev.oapd.inaf.it/cgi-bin/cmd}} span the stellar age interval from 4 Myr to 12.6 Gyr and the metallicity interval from $5 \times 10^{-3}$ to $1.6~Z_{\sun}$.
We neglect the effect of dust attenuation.

Then, the galaxy evolves nearly in isolation until $z\approx 6.5$, when it starts to interact with another galaxy in a mildly minor $q\approx5$ merger.
During this period, the total gas fraction $f_{\rm gas} = M_{\rm gas} / (M_{\star} + M_{\rm gas})$ within $0.1 R_{\rm vir}$ oscillates around 50\% (only mildly increasing with time), while the fraction of cold gas ($T < 10^4$~K) is lower and about 37\%.
The stellar mass, measured as the mass within $0.1 R_{\rm vir}$, is $M_{\star} \approx 2.5 \times 10^{9}$~M$_{\sun}$.
By changing the filtering scale (e.g. assuming a fixed volume of 3 physical kpc or 20 comoving kpc), the difference on the determination of $M_{\star}$ is less than 20\%.
During this phase, the global properties of the main galaxy, namely stellar mass, gas fraction, and star formation rate (see below), are quantitatively similar\footnote{\citet{watson+15} have observed a galaxy at $z = 7.5 \pm 0.2$ with stellar mass $1.7^{+0.7}_{-0.5} \times 10^{9}$~M$_{\sun}$, star formation rate $9^{+4}_{-2}$~M$_{\sun}$~yr$^{-1}$ (from infrared light), and gas fraction $55 \pm 25$ \%.} to those observed in a few, strongly lensed galaxies at $z \sim 7$ by \citet{bradley+12} and \citet{watson+15}.
Moreover, the stellar mass is a factor 2 higher than the halo mass-stellar mass relations determined by \citet{behroozi+13}, but yet consistent within the $2\sigma$ uncertainty.

Figure \ref{fig_SFH} shows the evolution of the star formation rate of the main galaxy as a function of time.
This is continuously increasing with time, with isolated peaks corresponding to the major mergers highlighted in Figure \ref{fig_merger_tree}.
Around $z \approx 6.5-7$, the main galaxy is forming $\sim 20$~M$_{\sun}$~yr$^{-1}$ in new stars.
On the other hand, the specific star formation rate decreases from $\gtrsim 10$~Gyr$^{-1}$ to $\sim 5-6$~Gyr$^{-1}$ at $z = 6.5$.
These numbers are consistent with recent determinations of the typical star formation rate of galaxies in samples at $z > 4$.
\citet{smit+12} have computed the star formation rate function (normalised as the number of galaxies per unit volume and unit star formation rate) starting from dust-corrected UV luminosity functions from \citet{bouwens+07,bouwens+11}.
They have found that between redshift 6 and 8 the star formation rate corresponding to the knee $L_{\star}$ of a Schechter luminosity function is about 10~M$_{\sun}$~yr$^{-1}$, suggesting that our galaxy is probably slightly larger than $L_{\star}$.
By correcting the value of $L_{\star}$ at $z=6.8$ determined by \citet{bouwens+11} as described by \citet{smit+12}, we get a corresponding absolute magnitude after dust correction $M_{\rm UV}^{\star} = -20.6$, which is indeed slightly larger (i.e. less bright) than the rest-frame magnitude $M_{\rm U} = -20.8$ of the stellar component within $0.1 R_{\rm vir}$ measured from run PH.
We caution however that the observational measurements have been obtained at 1600~\AA, while our estimate from the simulation comes from U rest-frame.
We also compare the specific star formation rate as a function of mass with a population of slightly lower redshift galaxies ($4.5< z<5.5$, with spectroscopic redshifts) observed in the \emph{VIMOS Ultra-Deep Survey} (VUDS) by \citet{tasca+15}.
At the mass of a few $10^9$~M$_{\sun}$, our galaxy sits slightly above the bulk of the data at $z \sim 5$ (thought there is large scattering), which in turn are distributed around the relation inferred by \citet{schreiber+15} from \emph{Herschel} data.
This is also consistent with the slowly growing trend of the average specific star formation rate with increasing redshift derived by \citet{tasca+15}, $\propto (1+z)^{1.2}$, which would predict $\sim 6$~Gyr$^{-1}$ at $z=6.5$, although most of the galaxies used to get this determination have $M_{\star} > 10^{10}$~M$_{\sun}$.
Overall, the global properties of the main galaxy in run PH seems to be consistent with typical star forming galaxies recently observed at similar redshifts \citep{iye+06,bradley+12,watson+15}.


\begin{figure}
\begin{center}
\includegraphics[width=\columnwidth]{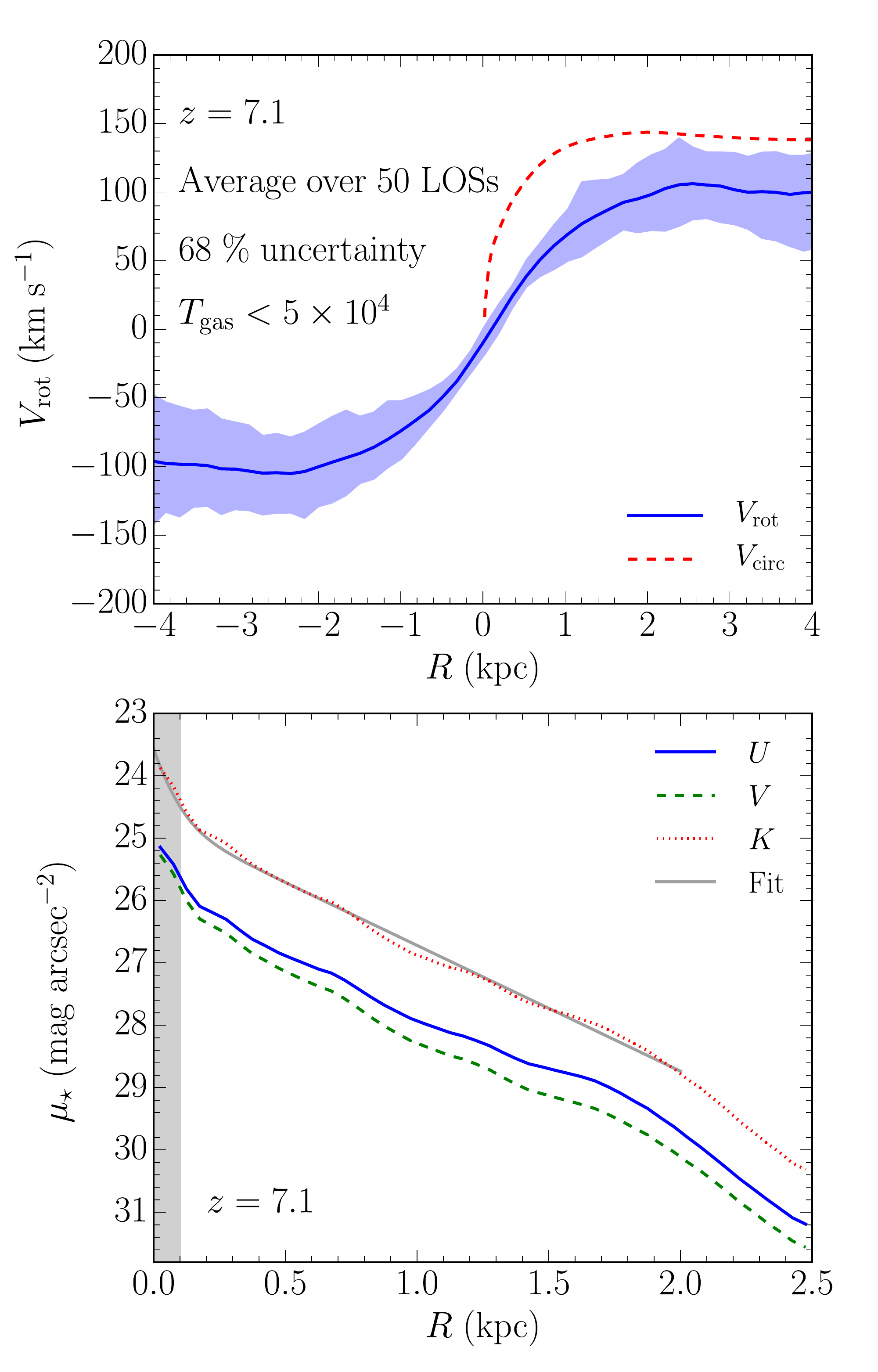}
\caption{Upper panel: rotation curve $V_{\rm rot}$ at $z = 7.1$ from the cold gas ($T < 50,000$~K).
The blue solid line is the average over 50 random line of sights through the galaxy mid-plane, while the shaded region marks the 68\% of the measurements.
The red dashed line is the circular velocity $V_{\rm circ} = \sqrt{G M(<r) / r}$ shown for comparison.
Lower panel: rest frame surface brightness profiles of the stellar component in U (blue solid), V (green dashed), and K (red dotted) bands at $z=7.1$.
The grey shaded region at the centre marks 2 gravitational softenings.
The surface brightness profiles include cosmological dimming.
The grey solid line shows the best-fit of the profile decomposition of $\mu_{\star}$ into two exponential profiles for the K band as an example.
}
\label{fig_rot_curve}
\end{center}
\end{figure}

\begin{figure*}
\begin{center}
\includegraphics[width=2\columnwidth]{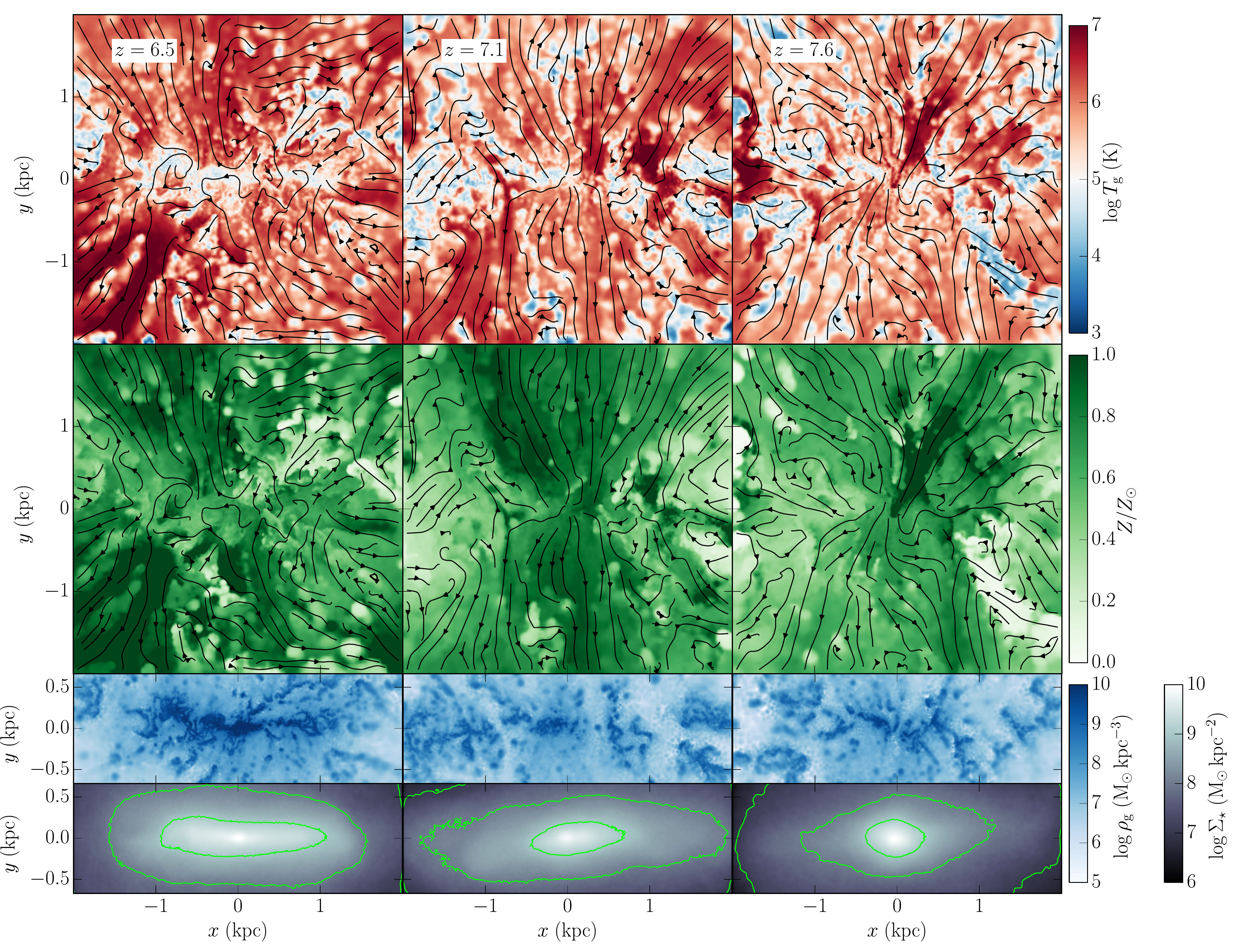}
\caption{From left to right: edge-on view of the main galactic disc in run PH between $z=7.6$ and $z=6.5$.
First row: gas temperature map averaged within a 200~pc thick slice perpendicular to the disc plane and centred on the disc centre.
The arrows show the velocity field of the gas.
Second row: the same of the top row, but for the gas metallicity.
Third row: gas density map averaged within a 200~pc thick slice as for the temperature and the metallicity.
Fourth row: stellar surface density projected within 4 kpc centred on the disc centre.
The green lines mark equal surface density contours corresponding to $10^{7}$, $10^{8}$ and $10^{9}$~M$_{\sun}$~kpc$^{-2}$, from outside inward.
}
\label{fig_disc_edgeon}
\end{center}
\end{figure*}

\subsection{Structure of the galactic disc}\label{sec_disc}

Figure \ref{fig_disc_faceon} shows the evolution of the galactic disc in run PH between $6.5 < z < 7.6$, after the last major merger is completed.
The galaxy is oriented face-on by determining the specific angular momentum of the gas within a sphere of radius $0.1 R_{\rm vir}$, after having centred the galaxy on the minimum of the potential and having removed the systemic velocity evaluated as the mass-weighted velocity of the particles within 500~physical~pc from the centre.
During this interval of time, the galaxy appears as a disc with tenuous nearly axisymmetric structures in the stellar component.
At $z=7.6$ the two cores originally at the centre of the merging galaxies are finally coalescing at the centre of the remnant, forming a tiny bulge (also visible as a redder component in Figure \ref{fig_merger_tree}).

The surface density map of the gas reveals that the gaseous disc is highly inhomogeneous, with many over-dense regions.
This is particularly evident at $z=6.5$, when the disc is less axisymmetric and more perturbed than at slightly larger redshifts, with more extended regions of dense gas at and above $\sim 10^8$~M$_{\sun}$~kpc$^{-2}$, possibly because of the interaction with the companion galaxy visible in Figure \ref{fig_merger_tree}.
Indeed, the gas is highly multi-phase, as also confirmed by the temperature maps.
Local fluctuations can vary from temperatures $< 1000$~K to $\sim 10^6$~K on scales of $\gtrsim 100$~pc.
Those high temperatures are likely triggered by local injection of energy from stellar feedback.

Such a structure of the interstellar medium is potentially a signature of turbulence, as expected from previous work on high-$z$ galaxies (e.g. \citealt{green+10,bournaud+11,hopkins+12}), though the main driver of turbulence is still debated.
Indeed, we observe large fluctuations in the local velocity dispersion of the gas, calculated as SPH kernel-weighted average of the local standard deviation of the velocity, ranging from a few to almost 100~km~s$^{-1}$, with a typical, average value $\sigma_{\rm g} \sim 40$~km~s$^{-1}$ across the disc at different redshifts.
We discuss the turbulence within the disc in more detail in Section \ref{sec_turbulence}.

High-density regions are naturally associated to larger star formation rates, following a large-scale Kennicutt-Schmidt relation \citep{kennicutt+98,bigiel+08} with slope $n \approx 1-1.5$ at (total) surface densities $> 1$~M$_{\sun}$~pc$^{-2}$, consistent with the local volumetric Schmidt law \citep{gnedin+14}.
The star formation within the disc is mostly distributed along the major features (i.e. spiral arms) in the stellar disc, with typical star formation rate surface densities $\gtrsim 1$~M$_{\sun}$~yr$^{-1}$~kpc$^{-2}$, while over-densities are associated with isolated regions of intense star formation as high as $\gtrsim 25$~M$_{\sun}$~yr$^{-1}$~kpc$^{-2}$.

Figure \ref{fig_rot_curve} shows the rotation curve $V_{\rm rot}$ and the surface brightness profiles of the main galaxy in run PH at $z = 7.1$.
The rotation curve is obtained as the mean over 50 random line of sights through the disc mid-plane.
We take into account only cold/warm gas with temperature $< 50,000$~K to mimic the rotation curve that could be obtained through HI/H$\alpha$ observations.
The galaxy has a flat rotation curve up to a few kpc from the centre with a nearly constant value about 100~km~s$^{-1}$.
The circular velocity $V_{\rm circ} = \sqrt{G M(<r) / r}$ is also flat, but with an asymptotic large velocity about 140~km~s$^{-1}$.
This mismatch witnesses the degree of non radial motion in the gas, also visible in the fluctuation of $V_{\rm rot}$ at $R \geq 2$~kpc. 
The surface brightness profiles of the stellar component in the rest frame U,V, and K band show that the stellar disc is nearly exponential, with a disc scale radius (half light radius) about $460$~physical~pc ($760$~physical~pc) in K band, which is grossly consistent with the size of typical galaxies at $z\sim 7$ \citep{oesch+10b}.
The stellar disc drops slightly outside 2~kpc and it shows a tiny steepening for radii below $\sim 200-300$~pc owing to the presence of a tiny bulge.
We perform a profile decomposition of the surface brightness in two exponential profiles, one for the disc and on for the bulge, finding a similar value of $B/T \approx 0.03\%$ (i.e. fraction of the total light in the bulge component) across the three photometric bands.

Figure \ref{fig_disc_edgeon} shows the edge-on evolution of the disc that forms in run PH.
The disc mid-plane corresponds to $y=0$, though it might look difficult to identify a smooth structure because the disc is indeed highly perturbed in the vertical direction, showing a clumpy and discontinuous structure.
We calculate the vertical density profile of the gas, azimuthally averaging the gas density in 5 linearly-spaced bins between 20~pc and 2~kpc from the centre of the disc.
We find that the gas density drops by almost an order of magnitude after a typical distance from the disc mid-plane between 100~pc and 300~pc, with the slight tendency to increase at larger cylindrical radii (i.e. the disc flares at large radii), as it is mildly hinted by Figure \ref{fig_disc_edgeon}.
Indeed, we fitted the vertical profile of the gas density with an exponential law, finding typical vertical scales $\approx 300$~pc.
A similar result (even a slightly ticker disc, probably because of the different coupling between feedback energy and dense gas) is obtained analysing the PH\_PF1 run with the reduced pressure floor, suggesting that the latter does not play a major role in setting the vertical structure of the disc.
We repeat the same analysis for the stellar disc and we find that the latter is somewhat thinner, with exponential vertical scales typically $\approx 170-200$~pc.
This is consistent with stars forming preferentially in the denser and thinner cold gas disc.

The vertical temperature structure shows interesting features.
The dense, clumpy gas that can be identified in the central disc by comparing with the density map is typically cold, with temperatures between $\gtrsim 100$ and $\sim 8000$~K owing to metal cooling, and it is embedded in an atmosphere of hot gas $\gtrsim 10^5$~K.
The hot gas is typically outflowing vertically from the central disc, funnelled through clumps of colder gas, likely suggesting that it originates from episodes of supernova feedback.
This gas is indeed polluted to metallicity $\lesssim Z_{\sun}$ and is injected into the circumgalactic medium.

\begin{figure}
\begin{center}
\includegraphics[width=\columnwidth]{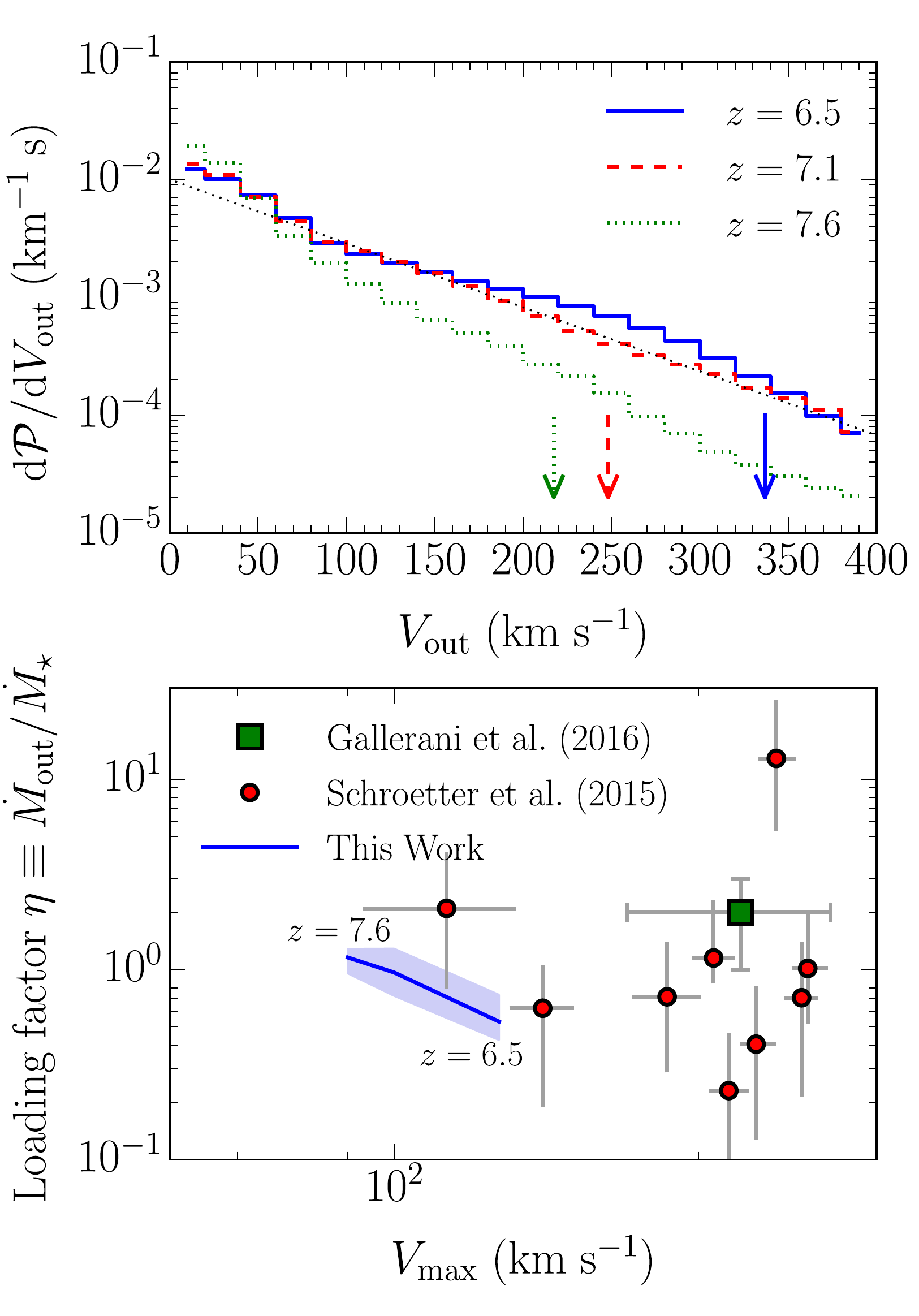}
\caption{Upper panel: mass-weighted probability distribution of the outflow velocity $V_{\rm out}$ at $z=6.5$ (blue solid line), $z=7.1$ (red dashed line), and $z=7.6$ (green dotted line).
The arrows show the escape velocities at each $z$.
The black dotted lines shows ${\rm d}\mathcal{P} / {\rm d}V_{\rm out} \propto \exp(-V_{\rm out}/\tilde{V})$, with $\tilde{V} = 80$~km~s$^{-1}$.
Lower panel: mass loading factor $\eta$ as a function of the asymptotic maximum rotation velocity.
The red circles are the observational data from \citet{schroetter+15} and references therein,
the green square is the average determination of \citet{gallerani+16} from the catalogue of $z \sim 5$ galaxies by \citet{capak+15},
and the blue line shows the results for run PH between $z=7.6$ and $z=6.5$.
The blue shaded area is spans the results associated to different shells used to measure $\dot{M}_{\rm out}$ (see the text for details).
}
\label{fig_outflows}
\end{center}
\end{figure}

Figure \ref{fig_outflows} shows the probability distribution of the outflow velocities $V_{\rm out}$ (i.e. the radial velocity $v_{r} > 0$ of outflowing gas) of the gas particles within a spherical shell between $0.2$ and $0.3~R_{\rm vir}$ from the central galaxy (e.g. \citealt{muratov+15}).
The particles are selected to have $\bmath{v} \cdot \hat{\bmath{r}} > 0$, where $\bmath{v}$ is the particle velocity (after removing the systemic velocity of the halo) and $\hat{\bmath{r}}$ is the versor along the direction from the centre of the halo to the particle position.
The distribution is roughly exponential, i.e. $\propto \exp(- V_{\rm out} / \tilde{V})$, where $\tilde{V} \approx 80$~km~s$^{-1}$ at $z= 7.1-6.5$ ($\approx 60$~km~s$^{-1}$ at $z= 7.6$) is a typical scale value for the positive radial velocity.
In particular, $\tilde{V} \approx \sigma_{r} \equiv \sqrt{\langle V_{\rm out}^2 \rangle - \langle V_{\rm out} \rangle^2}$ (where $\langle \cdot \rangle$ is intended as mass-weighted average), suggesting that the genuinely outflowing material populates the extended tails up to velocities as high as $200-300$~km~s$^{-1}$.
However, the fraction of the outflowing mass at high velocity, more specifically with $V_{\rm out}$ larger than the escape velocity $V_{\rm esc}$, is typically low, oscillating between 1 and 5\%, i.e.
most of the gas expelled from the disc will be recycled through the halo and will eventually join the central galaxy.
Here, we define the escape velocity $V_{\rm esc}$ at the position $\bmath{x}$ of a particle as $V_{\rm esc} = \sqrt{2 |\phi(\bmath{x}) - \phi_{0}|}$, where $\phi$ is the local gravitational potential, and $\phi_0$ is the reference potential at $R_{\rm vir}$ calculated as the average gravitational potential of all particles in a thin spherical shell between $0.95~R_{\rm vir}$ and $R_{\rm vir}$.
We have repeated this analysis within two additional spherical shells, (i) with an outer radius $0.2~R_{\rm vir}$ and a thickness of 1 physical kpc, and (ii) between 2 and 2.5 physical kpc, and we have found very similar results, within a factor 2.

We have also computed the mass loading factor $\eta \equiv \dot{M}_{\rm out} / \dot{M}_{\star}$, where $\dot{M}_{\rm out} = \Delta r^{-1} \sum_i m_{i} \bmath{v}_{i} \cdot \hat{\bmath{r}}_{i}$ is the mass outflow rate, the sum involves only gas particles $m_{i}$ within the considered shell with $\bmath{v}_{i} \cdot \hat{\bmath{r}}_{i} > 0$, and $\Delta r$ is the radial thickness of the shell.
The results between $z=7.6$ and $z=6.5$ are shown in the lower panel of Figure \ref{fig_outflows}, where we compare with the observational determinations of $\eta$ by \citet{schroetter+15} and references therein.
They are plotted against $V_{\rm max}$, i.e. the asymptotic maximum of the rotation curve \citep{schroetter+15}.
We repeat the same analysis as for Figure \ref{fig_rot_curve} to calculate $V_{\rm max}$ from the simulation data, whose ``uncertainty band'' in Figure \ref{fig_outflows} is associated to the measurements within different shells.
We also show the recent determination of $\eta \approx 2$ (where we add a generous uncertainty of 50\%) by \citet{gallerani+16}, which represents the average $\eta$ estimated for the sample of $z \sim 5$ and $M_{\star} \sim 10^{10}$~M$_{\sun}$ galaxies presented in \citet{capak+15}.
In this case, we estimate $V_{\rm max}$ by using the stellar mass-halo mass relation by \citet{behroozi+13} to determine the virial velocity of the typical halo where such galaxies are supposed to live in at $z = 5$, with an uncertainty of 50~km~s$^{-1}$.
We find $\eta \approx 0.5-1$ for PH, in fair agreement with the observational estimates.
We have also repeated the analysis selecting particles with $\bmath{v}_{i} \cdot \hat{\bmath{r}}_{i} > \sigma_{r}$ and finding slightly lower values for $\eta$ by at most 40\%.
We caution though that 
the observational data from \citet{schroetter+15} are a collection of measurements at much lower redshift than our simulation, namely between 0.1 and 0.8.
However, it is noteworthy that the observational estimate at $z \sim 5$ by \citet{gallerani+16} is yet consistent with the lower redshift data and in particular with local starbursts \citep{heckman+15}.
Moreover, the agreement at face values is promising, given the scattering of order $>10$ in $\eta$ among different successful theoretical/empirical models (see e.g. Fig. 10 of \citealt{schroetter+15}).

\begin{figure}
\begin{center}
\includegraphics[width=\columnwidth]{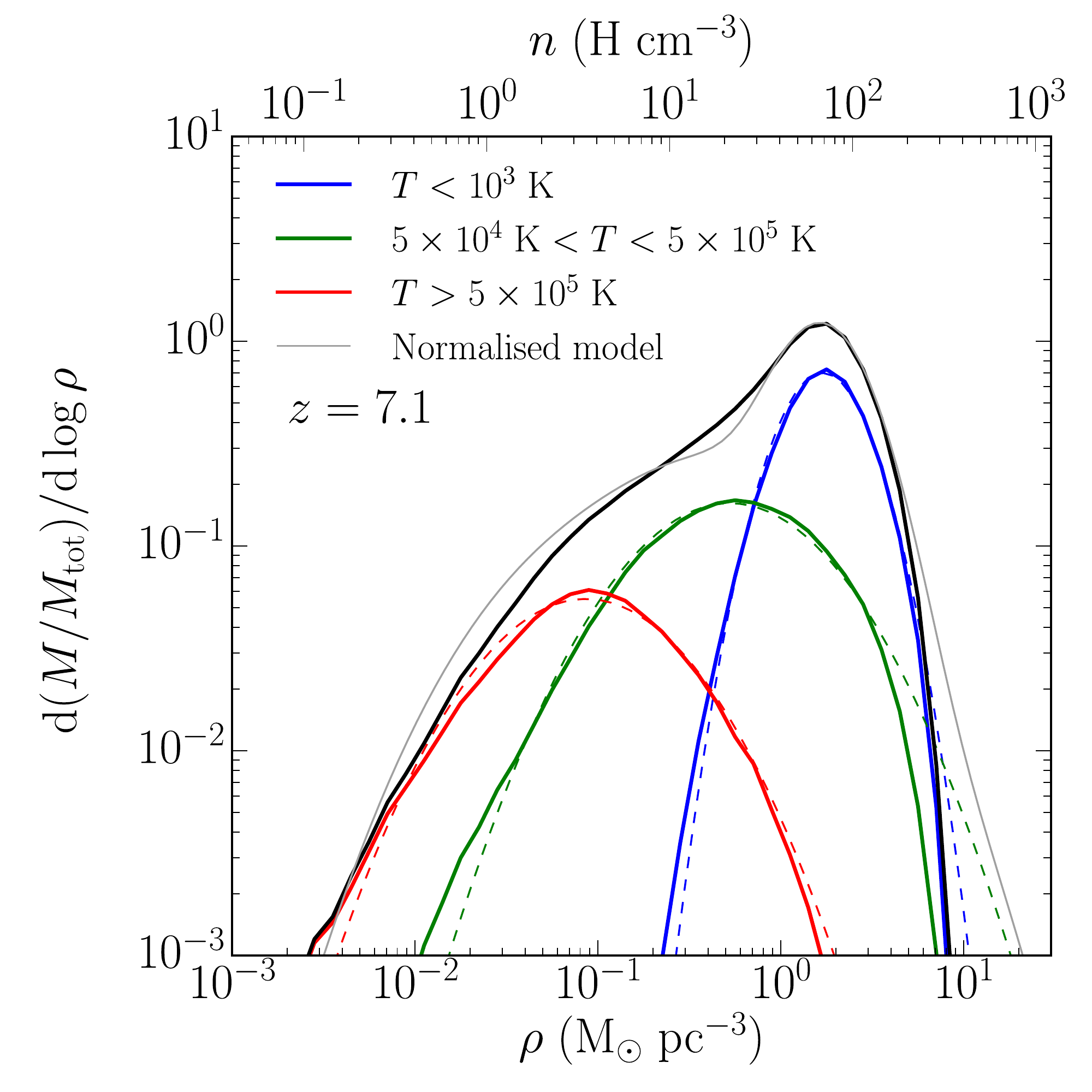}
\caption{Mass-weighted probability density function of the gas density within the galactic disc at $z=7.1$.
The black solid line shows the total distribution, while the blue, green, and read solid lines show the distribution of the ``cold'' ($T < 10^3$~K), ``warm'' ($5 \times 10^{4}~{\rm K} < T < 5 \times 10^{5}~{\rm K}$), and ``hot'' ($T > 10^{5}$~K) gas, respectively.
The dashed lines of with the same colour show the log-normal best fit of each component.
The thin grey line shows the compositions of the log-normal best fits, renormalised by the fractional mass occupied by the three selected components \citep{robertson+08}.
The fits describe fairly accurately the total density distribution.
}
\label{fig_den_pdf}
\end{center}
\end{figure}

On the other hand, clumps and streams of cold gas are typically raining down radially on the galactic disc.
Some of that gas is mildly polluted $Z \sim 0.2~Z_{\sun}$ and typically lives in over-dense regions, compressed by the surrounding hot outflows, where the cooling time is short, likely because of the metals already seeded by previous stellar outflows \citep{costa+15}.
However, some of the inflowing material has low metallicity, possibly coming from direct feeding through cold flows.
We select gas particles at $z=6.5$ within a cubic box 6~physical kpc in size centred on the main galaxy (removing a slab 400~pc thick centred on the disc mid-plane) with temperature $T < 5 \times 10^{4}$~K, i.e. the gas that is (or will be soon) able to form stars.
We trace them back among snapshots to $z=8$, storing the time-evolution of the density, temperature, and metallicity.
We select the gas that is directly inflowing from cold flows as the gas whose temperature never exceeds $10^{5.5}$~K, finding that it represents $\sim 50-60\%$ of the mass of cold gas originally selected in the trial volume.
A fraction $\sim 5-10\%$ of that gas is also accreted at nearly primordial composition as its metallicity always remains below $0.1~Z_{\sun}$.
We plan to devote a more detailed analysis to the gas inflow form larger scales as well as to the recycling of the gas within the halo in a forthcoming publication.


\subsection{The properties of the disc turbulence}\label{sec_turbulence}

The gaseous disc of the main galaxy is turbulent and multi-phase.
We show the mass-weighted probability density function of the gas within the disc at the representative redshift $z=7.1$ in Figure \ref{fig_den_pdf}.
We select the gas within a cylinder with a radius of 2~physical~kpc and 500~pc thick, centred on the disc mid-plane.
The density distribution is not approximately a log-normal distribution, as generally found in simulations of isothermal turbulence (e.g. \citealt{padoan+97,price+10}).
Instead, it peaks at a few M$_{\sun}$~pc$^{-3}$ ($\sim 100$~H~cm$^{-3}$), with an extended tail at low densities well below $10^{-2}$~M$_{\sun}$~pc$^{-3}$ ($\sim 0.2$~H~cm$^{-3}$) and a fast decline around $\sim 10$~M$_{\sun}$~pc$^{-3}$ ($\sim 250$~H~cm$^{-3}$).
This more complicated shape is a natural consequence of the mixing different phases of gas temperatures \citep{robertson+08}.
We can identify three main, nearly-isothermal components\footnote{In the following, we dub the different components within quotes in order to avoid confusion with the common phases of the interstellar medium \citep{ferriere+01}. In fact, their names are just labels for the temperature ranges used for the selection and, though similar to some of the phases of the interstellar medium, they do not refer directly to them.} characterised by an approximately log-normal distribution: a ``cold'' component at $T < 10^{3}$~K, a ``warm'' component with $5 \times 10^{4} < T/{\rm K} < 5 \times 10^{5}$, and a ``hot'' component a temperatures larger than $5 \times 10^{5}$~K.
They are associated with gas at average densities of 100, 10, and 1~H~cm$^{-3}$, respectively.
As a whole, they represent $65\%$ of the total gas in the disc, subdivided between ``cold', ``warm'', and ``hot'' as $\approx60 \%$, $\approx 29 \%$, and $\approx 11\%$, respectively.
The log-normal fits of the ``warm'' and ``cold'' components do not agree very well with the data at large densities.
This is possibly due to the effect of the pressure floor that limits the development of very high over-densities.
Indeed, repeating the same analysis on the run PH\_PF1 we find a lower peak of the ``cold'' component, which instead extends up to densities a factor $\sim 2-3$ larger, up to about 20~M$_{\sun}$~pc$^{-3}$.
In this case, the high density tail is nicely described by the log-normal distribution and it does not show a power law behaviour expected for highly self-gravitating turbulent flows (e.g. \citealt{scalo+98,federrath+08,kritsuk+11}), at least at the densities of a few hundreds H~cm$^{-3}$ that we are able to probe at the resolution of our simulation.

\begin{figure}
\begin{center}
\includegraphics[width=0.9\columnwidth]{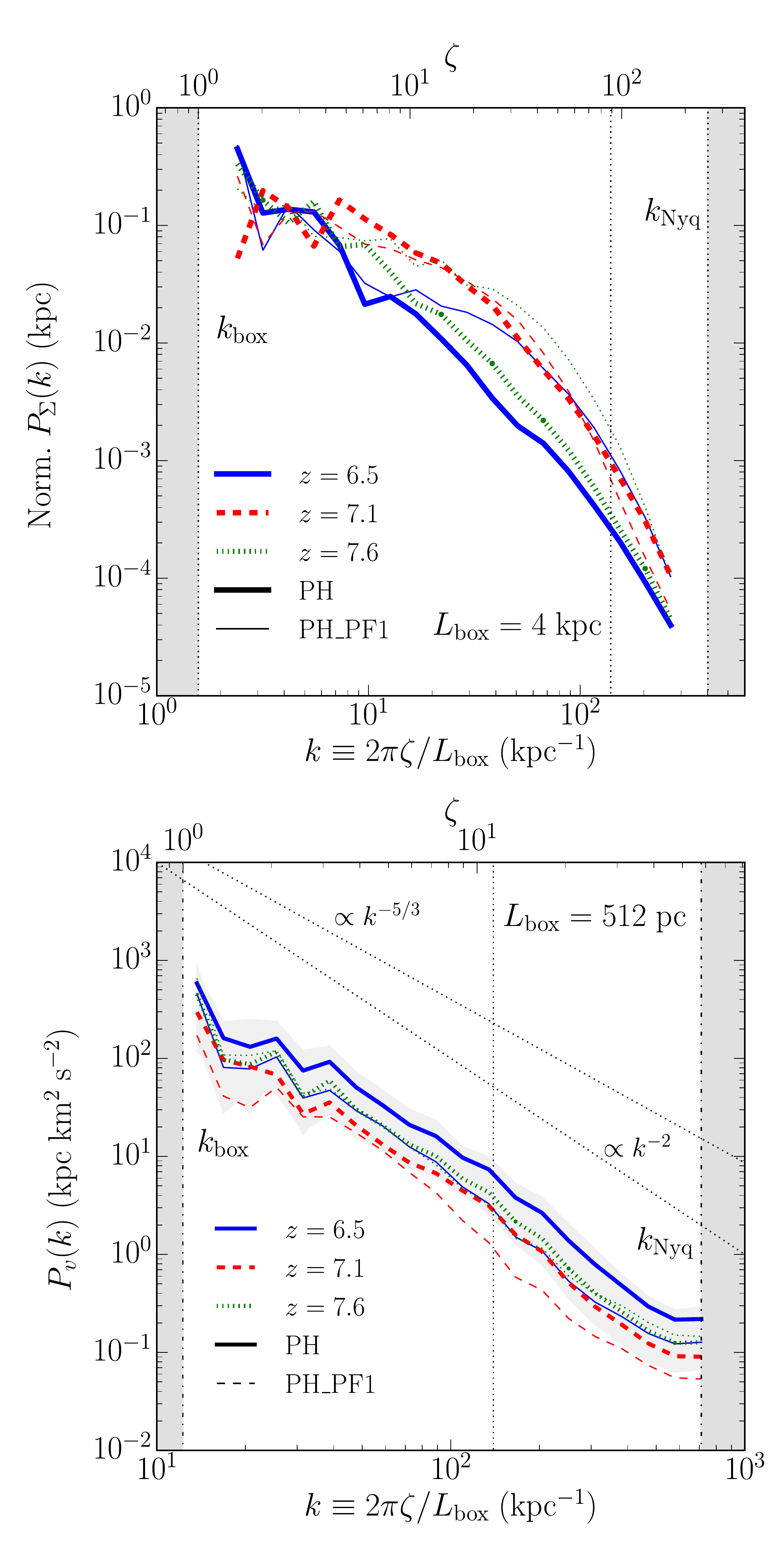}
\caption{Upper panel: 2-dimensional surface density power spectra of the gas in the galactic disc in a box $L_{\rm box} = 4$~kpc.
Blue solid, red dashed, and green dotted lines refer to $z=6.5$, 7.1, and 7.6, respectively.
Thick and thin lines are associated with run PH and PH\_PF1, respectively.
Lower panel: 3-dimensional velocity power spectra, obtained as average between 3 boxes $L_{\rm box} = 512$~pc (see the text for details).
The grey bands show the minimum-maximum range between the boxes at each redshift for run PH.
The line style is the same as above.
The dashed lines show $P_{v} \propto k^{-5/3}$ and $\propto k^{-2}$ for visual guidance.
In both panels, the vertical dotted lines mark the $k$ associated with the gravitational softening, while the grey shaded regions refers to $k < k_{\rm box}$ and $k > k_{\rm Nyq}$.
}
\label{fig_turbulence}
\end{center}
\end{figure}

We analyse the properties of the turbulence by computing the power spectrum of velocity and density
fluctuations at different scales.
Given any quantity $w$, its two-points correlation function within a volume $V$ is defined as:
\begin{equation}
\xi_{w}(\bmath{l}) = \frac{1}{V} \int_{V} w(\bmath{x} + \bmath{l})~w(\bmath{x})~{\rm d}^{3}\bmath{x},
\end{equation}
which can be generalised for a vectorial quantity by means of the inner product $\bmath{w}(\bmath{x}) \cdot \bmath{w}(\bmath{x}+\bmath{l})$.
The power spectrum of $w$ is the Fourier transform of $\xi_{w}$, which can be rearranged as:
\begin{equation}
p_{w} (\bmath{k}) = \frac{1}{(2 \pi)^{3/2} V} \left| \int_{V} w(\bmath{x})~e^{-i \bmath{k} \cdot \bmath{x}}~{\rm d}^{3}\bmath{x} \right|^{2}.
\end{equation}
We further assume that each mode is isotropic and it depends only on the module of the wavenumber $k = |\bmath{k}|$.
Then, the power spectrum becomes:
\begin{equation}
P_{w}(k) = k^{2} \int_{4 \pi} p_{w}(k, \theta, \phi) \sin\theta~{\rm d}\theta~{\rm d}\phi,
\end{equation}
where $(\theta, \phi)$ are the spherical coordinates in $\bmath{k}$-space.

\begin{figure*}
\begin{center}
\includegraphics[width=2\columnwidth]{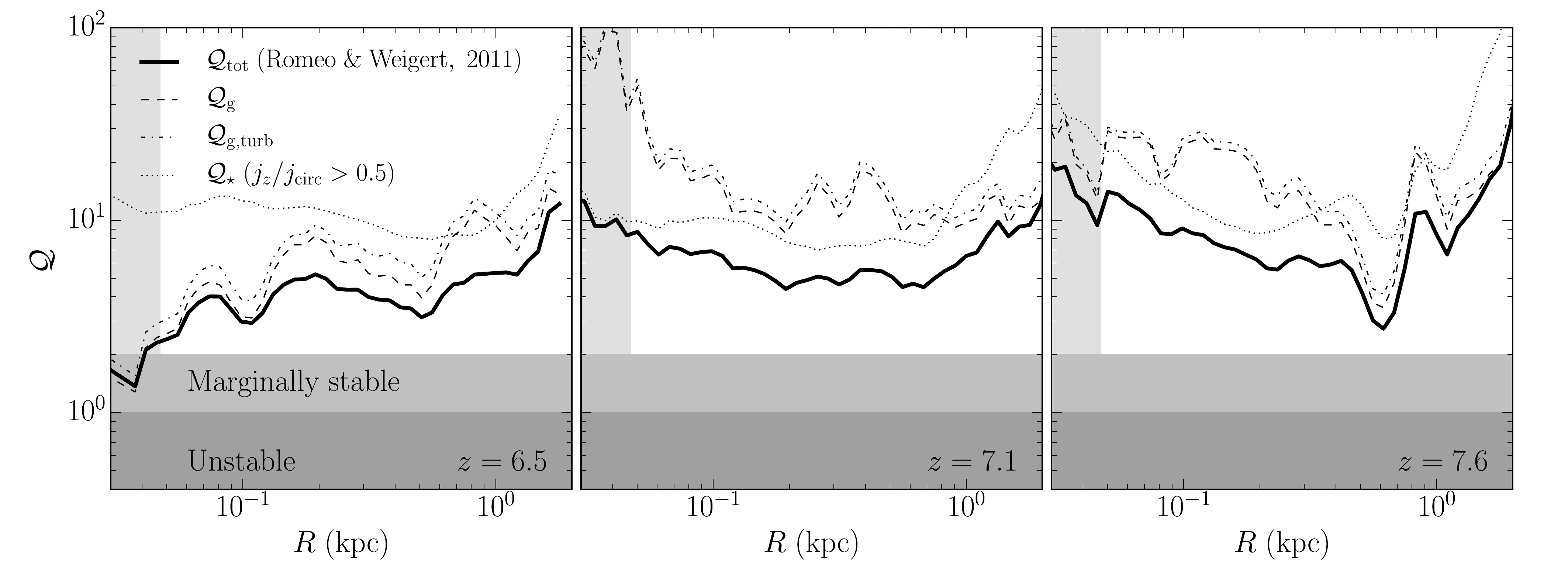}
\caption{Radial profiles of Toomre parameter $Q$ at $z=6.5$, 7.1, and 7.6.
Solid, dashed, dot-dashed, and dotted lines refer to the total $Q_{\rm tot}$ (according to \citealt{romeo+11}), the gas $Q_{\rm g}$ without the contribution of the gas velocity dispersion, the gas $Q_{\rm g,turb}$ including the gas velocity dispersion, and the stellar $Q_{\star}$ for stars in the disc with $j_{z} / j_{\rm circ} > 0.5$ (see the text for additional details).
The vertical grey region marks the gravitational softening, while the horizontal ones show the regions of instability and marginal stability.
}
\label{fig_qtoomre}
\end{center}
\end{figure*}

We perform this calculation for the gas surface density and for the velocity.
In the first case, we build a two-dimensional surface density map $\Sigma_{i, j}$ by projecting the SPH gas density field within 
a cube of $L_{\rm box} = 4$~physical~kpc per side centred on the main galaxy (oriented face-on as described in Section \ref{sec_disc}) on a grid 
of $512 \times 512$ square grids $\Delta x \approx 7.8$~pc per side, similar to the typical smoothing length within the disc, i.e. $\sim 10$~pc.
Then, we use the Fast Fourier Transform to numerically calculate the two-dimensional power spectrum $P_{\Sigma}$, after 
having padded with zeros the boundaries of the grid in order to reduce the aliasing in Fourier space produced by the 
non-periodic and finitely sampled content of the grid.

In the second case, we analyse the velocity fluctuations of the interstellar medium in the disc.
First, we selected three boxes within the disc with side $L_{\rm box} = 512$~physical~pc at 1.5~kpc from the centre and at 120$\degr$ angular separation from each other, in order to avoid any correlation among them and to minimise the effect of differential rotation since the rotation curve is roughly flat outside the central kpc (see Figure \ref{fig_rot_curve}).
Each box typically contains from at least a few tens of thousands to a few hundreds of thousands gas particles.
We subtract the systemic velocity of each cube as the mass-weighted gas velocity and we then interpolate the SPH-averaged gas velocity on a cubic grid of $128\times 128\times 128$ using the {\sc tipgrid} code\footnote{{\sc tipgrid} was written by Joachim Stadel and it is available at \url{http://astrosim.net/code/doku.php?id=home:code:analysistools:misctools}.}.
Then, we finally proceed as described above to calculate the velocity power spectrum $P_{v}$. 
The velocity power spectrum $P_{v}$ is normalised such as its integral over $k$ gives the total velocity dispersion $\sigma_{\rm g}^2$ of the gas in each box, while $P_{\Sigma}$ is normalised such as its integral over $k$ gives 1.
In both cases, we compute the power spectrum between $k_{\rm box} = 2 \pi / L_{\rm box}$ and the Nyquist wavenumber $k_{\rm Nyq} = \pi / \Delta x$.

The results of these calculations are shown in Figure \ref{fig_turbulence}.
The power spectrum of the surface density grossly shows a two power laws behaviour, with a break at $k \sim 30$~kpc$^{-1}$, which corresponds to a scale length $\sim 200$~pc.
This is roughly consistent with our previous estimate of the disc scale height, as generally found in simulations \citep{bournaud+10} as well as observations \citep{elmegreen+01}.
This suggests a transition between more two- and three-dimensional like turbulence at low and high $k$ (or large and small length scales), respectively.
The power law exponents oscillate significantly with redshift, ranging roughly between -1.0 and -1.6 for low $k$ and between -2.3 and -2.7 for high $k$, which in both cases is slightly less than what \citet{bournaud+10} found.
Comparing the results with the run PH\_PF1, we find similar scatterings, but marginally different exponents (note however that the power spectra at $z=7.1$ are fairly similar among the two runs).
Typically, the large scale modes have slightly shallower slopes between -0.6 and -1.2, while the power at small scales decays faster with $k$, typically having steeper slopes around -3.
Moreover, the transition between the low- and  high-$k$ branch and the latter as well are at somewhat higher normalisation compared with the reference run.
This different behaviours imply that there is slightly more power in large scale modes and at the transition between two- and three-dimensional modes at smaller scales than in the reference simulation, likely because the lower pressure support promotes slightly larger over-densities to develop under the influence of self-gravity, though the gross structure of the disc (i.e. the disc thickness) roughly remains the same.

At globally smaller scales, the velocity power spectra (lower panel of Figure \ref{fig_turbulence}) show much less differences in shape among the runs with different pressure floors, i.e. runs PH and PH\_PF1, though there are fluctuations of factors $\sim 2-3$ in the normalisations at different redshifts.
This suggests that the development of turbulence is only mildly influenced by the presence of the pressure floor at any scale larger than the gravitational softening.
The typical power law slopes measured between $k_{\rm box}$ and $k = 100$~kpc$^{-1}$ (i.e. at scales larger than the gravitational softening) is about 1.6 with little variations among runs and redshifts.
This looks consistent with a Kolmogorov-like turbulence spectrum, i.e. for incompressible, subsonic turbulence; however, when we remove from the fits the first 2 modes, which are poorly sampled due to the box size, we obtain a slope close to -2, again with little scattering among runs and redshifts.
Such a slope matches the Burgers law expected for compressible supersonic turbulence (e.g. \citealt{federrath+13}).
However, it is difficult to unequivocally discriminate between the two descriptions; 
in fact, the gas is a mixture of phases characterised by different temperatures and Mach numbers since the density distribution of each box is similar to what shown in Figure \ref{fig_den_pdf}.
We typically find a transonic or mildly supersonic turbulence with Mach number $\sigma_{\rm g} / \langle c_{\rm s} \rangle \gtrsim 1$, where $\langle c_{\rm s} \rangle$ is the mean sound speed determined from the mass-weighted average temperature of the gas (i.e. averaged over the different phases) about $10^5$~K.
Nonetheless, we argue that the gas within the disc might be slightly better described by a Burgers-like law because of its  compressible nature and the larger mass fraction in cold and supersonic gas.


\begin{figure*}
\begin{center}
\includegraphics[width=2\columnwidth]{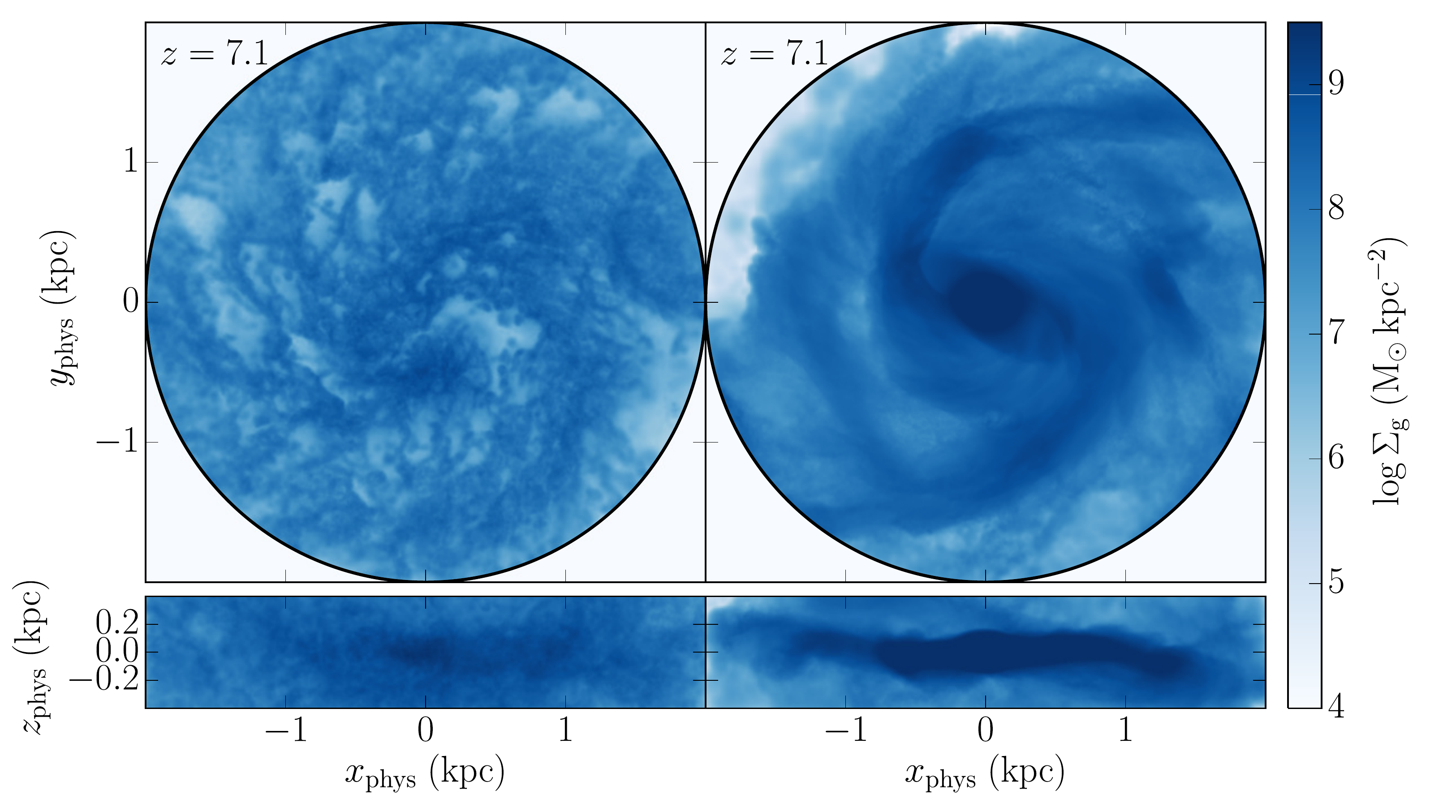}
\caption{Gas surface density maps of the gaseous disc of the main galaxy at $z=7.1$ in runs PH (left column) and PH\_NF (right column). 
The projections are face-on (upper row) and edge-on (lower row); sizes are in physical units.}
\label{fig_disc_nofb}
\end{center}
\end{figure*}

\subsection{What shapes the interstellar medium at high redshift?} \label{sec_ism}

Several physical processes may contribute to fostering turbulence in the interstellar medium.
Among them, gravity and stellar feedback are often considered to be the dominant ones \citep[e.g.][]{gomez+02,joung+06,brunt+09,pan+15}.
Gravity can trigger turbulent motions in a disc -- gravito-turbulence -- when the local cooling time is factor of few to several the local orbital time.
Despite that the exact transition is still matter of debate \citep{gammie+01,meru+11a,meru+11b,paardekooper+11,lodato+11}, a gravito-turbulent state is typically characterised by a Toomre parameter $Q \approx 1.5-2$.
Figure \ref{fig_qtoomre} shows the Toomre parameter of the disc in run PH at different redshifts.
We measure the Toomre parameter within a disc 2.5 physical kpc in radius and 500 physical pc thick.
The general definition of the Toomre parameter is:
\begin{equation}\label{eq_toomre_q}
Q = \frac{\kappa V}{A G \Sigma},
\end{equation}
where $\kappa = \sqrt{2 (V_{\phi}/R)^2 (1 + {\rm d}\log V_{\phi} / {\rm d}\log R)}$ is the epicyclic frequency defined through the azimuthal velocity $V_{\phi}$ and the polar radius $R$, and $\Sigma$ is the surface density (either of the gas or the stars).
The factor $A$ is $A_{\rm g} = \pi$ and $A_{\star} = 3.36$ for gas and stars, respectively.
The velocity $V$ is the radial velocity dispersion $V = \sigma_{R}$ for stars, and the sound speed $V = c_{\rm s}$ for gas.
If the gas is turbulent with a radial velocity dispersion $\sigma_{{\rm g,} R}$, the ``turbulent'' gas Toomre parameter $Q_{\rm g, turb}$ adopts the corrected velocity $V = \sqrt{c_{\rm s}^2 + \sigma_{{\rm g,} R}^2}$.

We correct the Toomre parameter given by equation (\ref{eq_toomre_q}) for finite disc thickness effects following \citet[][see also \citealt{romeo+94,romeo+13,inoue+16}]{romeo+11}; we multiply $Q$ by:
\begin{equation}
T = \left\{
\begin{array}{lc}
1 + 0.6 (\sigma_{z} / \sigma_{R})^2 & \sigma_{z}/\sigma_{R} < 1/2 \\
0.8 + 0.7 (\sigma_{z} / \sigma_{R}) & \sigma_{z}/\sigma_{R} \geq 1/2 \\
\end{array},
\right.
\end{equation}
where $\sigma_{z}$ and $\sigma_{R}$ are respectively the vertical and (polar) radial velocity dispersion, either of gas or stars.
We generally measure the velocity dispersions as $\sigma = \sqrt{\langle v^2 \rangle - \langle v \rangle^2}$, where $\langle \cdot \rangle$ is the SPH average on the smoothing kernel of the appropriate velocity $v$.

The Toomre parameter describes the stability of rotating discs.
However, the stellar component has a tiny bulge at the centre, which is dispersion-dominated.
Therefore, we calculate the ratio $\epsilon = j_z / j_{\rm circ}$, where $j_{z}$ is the $z$ component of the specific angular momentum of a particle computed in a reference frame centred on the galaxy, after having aligned the $xy$ plane with the galactic mid-plane.
The maximum angular momentum of a particle at distance $r$ from the centre is the angular momentum on a circular orbit, namely $j_{\rm circ} = r V_{\rm circ}(r)$, where we estimate the circular velocity $V_{\rm circ} \approx \sqrt{G M(<r) / r}$.
In order to identify the stars that belong to the rotating disc, we select those with $\epsilon > 0.5$, i.e. the stars whose angular momentum perpendicular to the disc plane is close to maximal.
Finally, we calculate the total Toomre parameter as:
\begin{equation}
Q_{\rm tot}^{-1} = \left\{
\begin{array}{lc}
W Q_{\star}^{-1} + Q_{\rm g}^{-1} & Q_{\star} \geq Q_{\rm g} \\
Q_{\star}^{-1} + W Q_{\rm g}^{-1} & Q_{\star} < Q_{\rm g} \\
\end{array},
\right.
\end{equation}
where $W = 2 V_{\star} V_{\rm g} / (V_{\star}^2 + V_{\rm g}^2)$, where $V_{\rm g}$ and $V_{\star}$ stand for the velocity adopted in equation (\ref{eq_toomre_q}) for the gas and the stars, respectively \citep{romeo+11}.

Figure \ref{fig_qtoomre} shows that the galactic disc is overall stable to gravitational perturbations, with the Toomre parameter typically ranging from $\sim 4$ to $\gtrsim 10$ across the redshift interval $z= 6.5-7.6$, i.e. after the disc rebuilt.
This is true for both the stellar and the gas components alone, and also when they are combined in $Q_{\rm tot}$, though the latter is slightly lower than the singular components.
Such a high value for $Q_{\rm g, turb}$ likely indicates that the gaseous disc is not in a gravito-turbulent state.
The only exception is at $z=6.5$, when the Toomre parameter of the gas component approaches 1.5-2 close to the centre, which in fact corresponds with a clumpy region in the gas, where over-densities might have been tidally enhanced by the close passage of the satellite galaxy that will eventually merger with the main one.
We have checked that $Q_{\star}$ goes to $\lesssim 1$ within the inner 100-200~physical~pc if we do not select particles with $\epsilon > 0.5$, that is kinematically consistent with the presence of a central tiny bulge.

Figure \ref{fig_qtoomre} also compares the Toomre parameter of the gas with and without the contribution of turbulence (i.e. $Q_{\rm g, turb}$ and $Q_{\rm g}$, respectively).
When the effect of turbulent motions is include, the Toomre parameter is naturally larger.
Nonetheless, such effect accounts for an increase in $Q$ up to a factor $\sim 2$.
This implies that the pressure support alone would be enough to prevent widespread fragmentation in the disc, because the average temperature $\sim 10^{5}$~K within the disc is such that the average sound speed of the gas is $\langle c_{\rm s} \rangle \sim \sigma_{\rm g} \sim 50$~km~s$^{-1}$, i.e. pressure support and non-thermal turbulence contributes similarly to the stability of the gaseous disc.

\begin{figure}
\begin{center}
\includegraphics[width=\columnwidth]{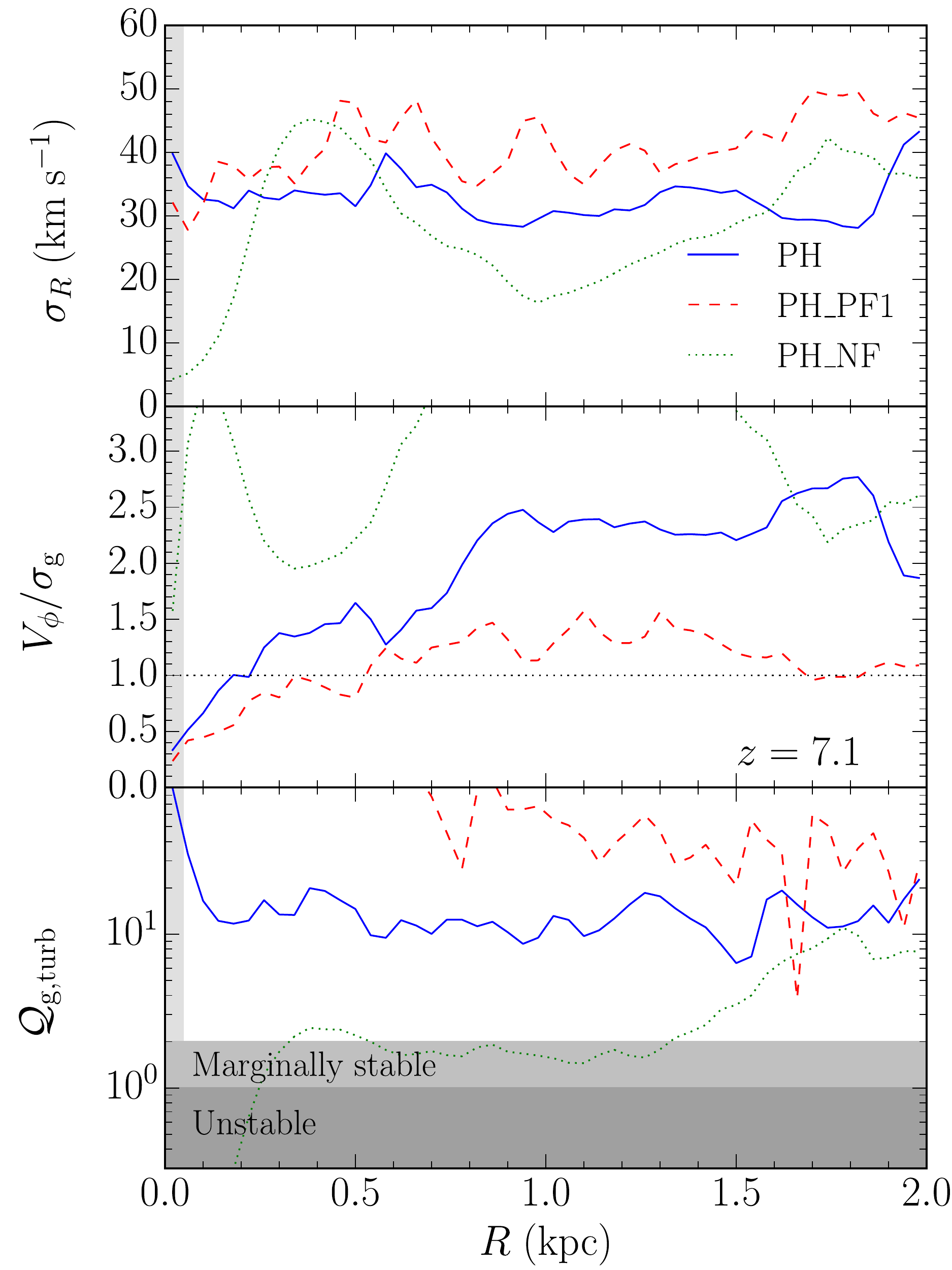}
\caption{Comparison between the disc properties in runs PH (blue solid line), PH\_PF1 (red dashed line), and PH\_NF (green dotted line).
From top to bottom: radial velocity dispersion of the gas, $V_{\phi} / \sigma_{\rm g}$ ratio of the gas, and turbulent Toomre parameter $Q_{\rm g, turb}$ of the gas.
The vertical grey region common to all panels marks the gravitational softening of the gas.
}
\label{fig_vdisp_nofb}
\end{center}
\end{figure}

The high value of the gas Toomre parameter and the mild sensitivity of the latter to turbulence suggest that (i) feedback may have a major role in regulating the stability of the disc, and (ii) disc might not be in a gravito-turbulent state, i.e. feedback is also the main energy source for the gas turbulence.
However, the latter point is hard to demonstrate because gravity can still contribute by accelerating free falling cold clouds from the surroundings that eventually dissipate their kinetic energy by mixing with the denser gas of the disc.
In order to test the impact of feedback on the structure of the interstellar medium, we have restarted the run PH without star formation nor feedback from $z\approx 8$, i.e. run PH\_NF.
Figure \ref{fig_disc_nofb} shows the comparison between the gaseous disc at $z=7.1$ in runs PH and PH\_NF.
The two discs are visibly different.
Under the effect of feedback, the disc in run PH is overall slightly more extended in radius and it has a lower surface density on average.
The gas in run PH has larger density contrasts on small scales and a more flocculent structure than in run PH\_NF, which on the other hand shows a clear two-armed spiral, initially triggered by the quasi-resonant tidal interaction with a flying-by satellite \citep[e.g.][]{donghia+10}. 
The disc in run PH\_NF is also thinner than in run PH and it presents a central bulge-like structure that is not dispersed by star formation and feedback and that is also mildly elongated in a bar-like fashion.

Figure \ref{fig_vdisp_nofb} tries to quantify the morphological differences between the results of runs PH and PH\_NF (comparing also run PH\_PF1) at $z=7.1$.
We compare the radial velocity dispersion of the gas $\sigma_{R}$, the ratio $V_{\phi}/\sigma_{\rm g}$ between the gas azimuthal velocity $V_{\phi}$ and the three-dimensional velocity dispersion $\sigma_{\rm g}$ ($V_{\phi}/\sigma_{\rm g}$ quantifies the rotational support of the gaseous disc), and $Q_{\rm g, turb}$.
We note that the gas in run PH\_NF tends to have a slightly lower
$\sigma_{R} \sim 25$~km~s$^{-1}$ than in both runs PH and PH\_PF1, except for a ``bump'' at $\sim 300-500$~pc where the rotating disc joins the central spheroid.
Below this radius, $\sigma_{R}$ drops to a few km~s$^{-1}$ in run PH\_NF, while it remains almost constant in the other cases, likely because the central gaseous bulge in run PH\_NF is mostly pressure supported.
This suggests an overall similar amount of turbulence with and without stellar feedback\footnote{The similarity of gas velocity dispersion achieved with and without feedback in presence of gravito-turbulence was pointed out in \citet{agertz+09}.}; however, other dynamical properties of the disc are very different among the runs.
Without stellar feedback, the disc is thinner and denser,  and consistently more rotationally supported than in the runs including feedback, as shown by $V_{\phi} / \sigma_{\rm g} \sim 3$, almost a factor of 2 larger than in the other cases.
Moreover, while $Q_{\rm g, turb} \sim 10$ is almost constant across the disc in run PH, it drops to $\lesssim 2$ within $R \sim 1.5$~kpc when the feedback is absent.
This is due to the combined effect of the larger surface density of the disc and of the lower local support provided by both turbulence and thermal pressure, since the gas temperature across the disc typically corresponds to $c_{\rm s} \sim 10$~km~s$^{-1}$ or lower when feedback is not included.
Within the central 500~pc (i.e. the gaseous bulge), $Q_{\rm g, turb}$ drops well below unity in run PH\_NF, but it loses its physical significance because the central region is not rotationally supported.

The gas velocity dispersions are similar in the case with and without feedback but the stability properties of the discs are not.
This suggests that a turbulent state with similar amplitude of turbulent motions can be achieved in different ways.
When feedback is active, the disc settles in a ``hot and turbulent'' configuration with a high Toomre parameter that results from efficient gas heating inside and beyond the disc.
Alternatively, when radiative cooling effectively counterbalances heating, the disc remains in a marginally unstable ``cold and turbulent'' state.
While all these pieces of evidence do not unambiguously disentangle the relevance of stellar feedback and gravity as sources of turbulence, they do show that stellar feedback has a major impact in shaping the interstellar medium of such high redshift galaxies.
We argue that stellar feedback is possibly the main source of turbulent energy in this case.
Indeed, only when we turn off stellar feedback the galaxy quickly readjusts to a new state that looks very similar to gravito-turbulence in terms of both morphology and structure of the disc, whereas when the feedback is active, the gaseous disc remains globally Toomre stable on a longer timescale.
This implies that supernova feedback affects the interstellar medium enough to prevent gravito-turbulence to take place and in turn it suggests indirectly that feedback mainly powers the turbulent cascade in the gaseous disc.

However, we caution that the interpretation of the differences between the runs with and without feedback is complicated by other aspects: 
(i) the PH\_NF run is restarted when the gas has already a notable amount of turbulent motions, hence its new dynamical state does not necessarily reflect only the onset of gravito-turbulence; (ii) the galaxy is still accreting both fresh material from larger scales and material recycled through the gaseous halo that likely contributes to additional turbulent motions when it joins the galactic disc \citep{klessen+10}. 
Regarding (i), though, it is noteworthy that the gas cools on timescales shorter than the orbital times when feedback is not active, which suggests that gravito-turbulence is needed to sustain the velocity dispersion in this new ``cold'' state of the disc.
In addition to that, the lack of stellar feedback makes PH\_NF grow more than PH owing to the lower ``resistance'' to gas inflows.
As a consequence, the baryonic mass (in particular the gas component) within $0.1 R_{\rm vir}$ of PH\_NF is $\sim 2$ times larger than in PH, which leads to a steepening of the circular velocity curve within $\sim 1.5$~kpc; the latter peaks at $\sim 190$~km~s$^{-1}$ and then asymptotically declines to $\sim 160$~km~s$^{-1}$.
This possibly stabilises the disc as it increases $\kappa$ in the numerator of equation (\ref{eq_toomre_q}).
At the same time, however, the lack of feedback increases the gas disc surface density through the enhancement of inflows and lowers the value of $V$, potentially decreasing $Q$.
Since we observe a lower value of $Q$ in PH\_NF, this suggests that feedback mostly influences the structure of the disc by means of both internal and external effects, namely the contribution to local turbulence and the indirect redistribution of mass through the enhancement of inflows, respectively.
All these evidences are consistent with the possibility that feedback has a prominent role in shaping the disc turbulence in our simulations; nonetheless, this complex interplay between feedback and gravity prevents a clear identification of the ultimate cause of the turbulent cascade in the interstellar medium.

Finally, we mention that, as described in Section \ref{sec_turbulence}, we have also computed the power spectrum of the density-weighted velocity $w = \rho^{1/3} v$ in run PH in order to distinguish between a solenoidal or compressive mode of energy injection in turbulence.
We find slopes typically $\sim -1.8$ that would favour a dominant solenoidal mode, according to the results of \citet{federrath+13}.
This looks consistent with recent work showing that supernova-driven turbulence is only mildly compressive, at least on the molecular cloud scale \citep{pan+15}, and it would support the interpretation that stellar feedback may be the main responsible for turbulence, though it is not clear whether and how this estimate is degenerate with and sensitive to the multi-phase structure of the gas.


\subsection{Mass flow through the disc}

\begin{figure}
\begin{center}
\includegraphics[width=\columnwidth]{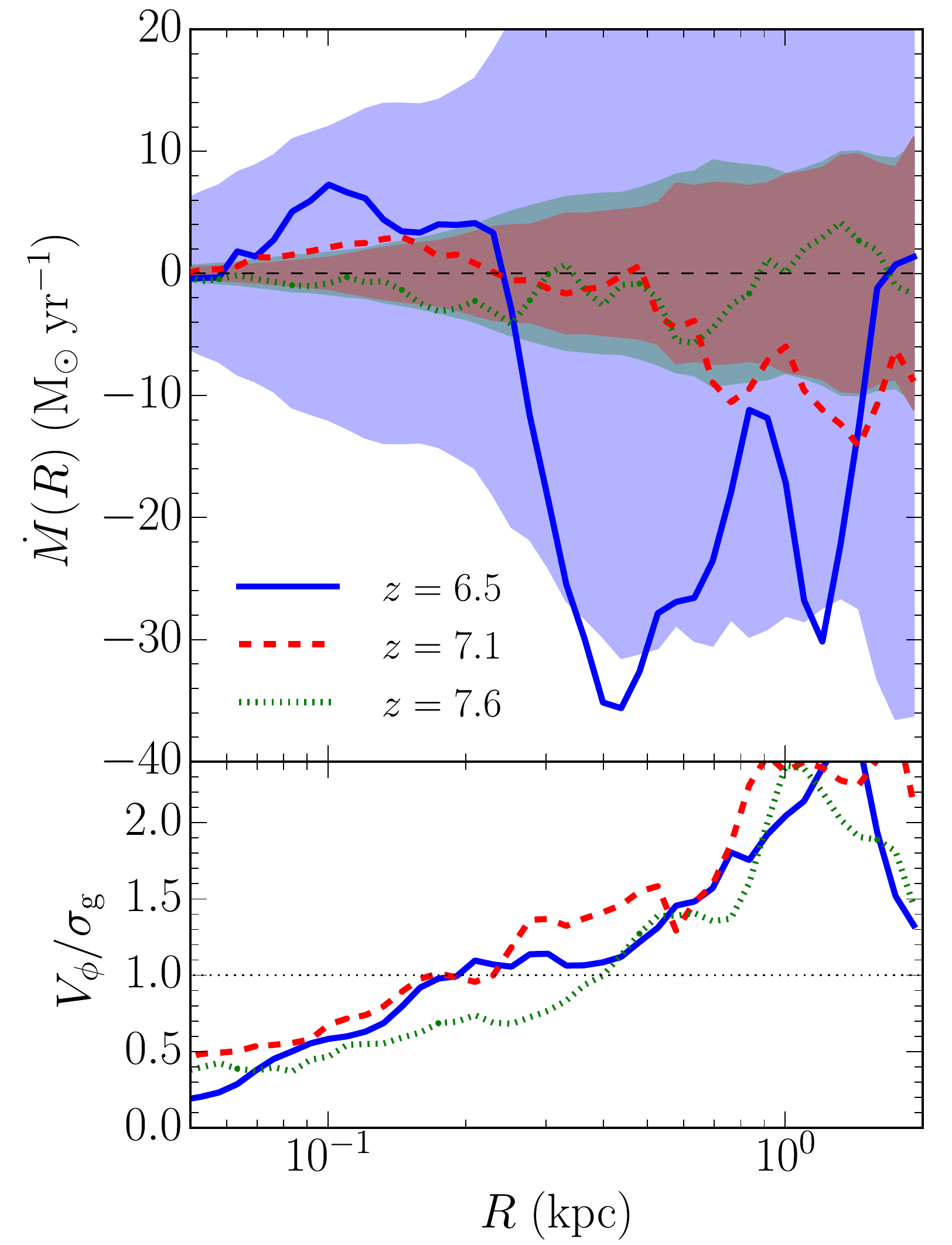}
\caption{Upper panel: radial mass flow through the gaseous disc.
Blue solid, red dashed, and green dotted lines refer to $z=6.5$, 7.1, and 7.6, respectively.
The shaded regions with the same colours mark the symmetric inflow-outflow estimated by equation (\ref{eq_mdot}).
Negative values of $\dot{M}$ are associated to inflows.
Lower panel: radial profile of $V_{\phi} / \sigma_{\rm g}$ at different redshifts.
}
\label{fig_mdot_profile}
\end{center}
\end{figure}

The turbulence within the disc can cause motions and mass transfer through the galactic disc.
The upper panel of Figure \ref{fig_mdot_profile} shows the radial mass flow through the galactic disc.
This is measured within a disc of 2 kpc radius and 400 pc thickness centred on the disc mid-plane.
For each radial bin of width $\Delta R$ in polar coordinates, we measure the mass flow as $\dot{M} = \Delta R^{-1} \sum_{j} m_{j} v_{j,R}$, where $m_j$ and $v_{j,R}$ are the mass and the radial mid-plane velocity of the $j$-th gas particle, respectively, and the sum extends over the particles within each radial bin.
Negative values of $\dot{M}$ are associated to inflows of gas through the galactic disc.

\begin{figure*}
\begin{center}
\includegraphics[width=2\columnwidth]{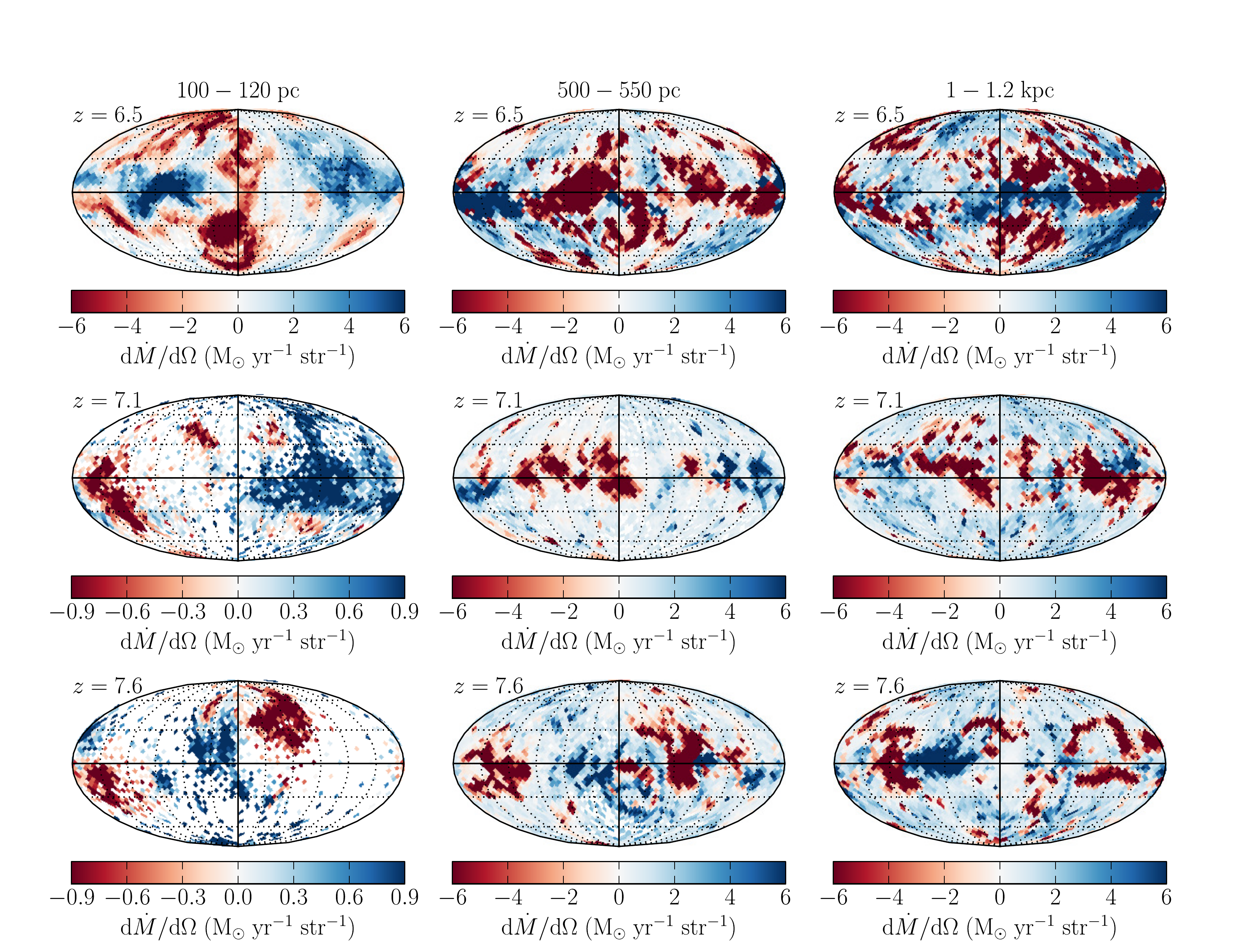}
\caption{Mollweide projections of the radial mass flow rate per unit steradian (negative and positive values represent inflow and outflow, respectively) at different scales and times.
From top to bottom: redshift 6.5, 7.1, and 7.6, respectively.
From left to right: mass flow rate in a spherical shell between 100 and 120 physical pc, between 500 and 550 physical pc, and between 1 and 1.2 physical kpc.
The equator of the map corresponds with the disc mid-plane.
}
\label{fig_mdot_map}
\end{center}
\end{figure*}

The figure shows that the gas is not steady inflowing toward the central region over the time span of $\sim 160$~Myr between $z=7.6$ and $z=6.5$.
Instead, $\dot{M}$ fluctuates from negative to positive values (i.e. from inflow to outflow) at different locations in the disc with typical absolute values within $\sim 10$~M$_{\sun}$~yr$^{-1}$.
However, the gas flows in more strongly at $z = 6.5$, up to about 30~M$_{\sun}$~yr$^{-1}$ between $R \approx 300$~pc and $\approx 2$~kpc, possibly because of the disturbance of the satellite galaxy visible in Figure \ref{fig_merger_tree}.
We can fairly estimate the absolute value of $\dot{M}$ over the disc as:
\begin{equation} \label{eq_mdot}
\dot{M} \sim M(<R) / t_{\rm turb} \sim M(<R) \sigma_{R} / R,
\end{equation}
where $M(<R)$ is the enclosed mass and $t_{\rm turb} \sim R / \sigma_R$ is the turbulence crossing time, which also approximates the dissipation time of the turbulent kinetic energy if not continuously replenished \citep{maclow+99,elmegreen+00}.
However, it does not constrain the sign of $\dot{M}$, i.e. whether gas would preferentially inflow or outflow.

The overall behaviour seems qualitatively consistent with mass transport due to turbulence induced by feedback and not gravito-turbulence, as discussed above.
Indeed, \citet{goldbaum+15} have used controlled simulations of gravitationally unstable disc galaxies to show that gravito-turbulence would be able to sustain a net mass inflow over time through the disc plane because of the coherent torquing of the gas from persistent spiral arms.
Specifically, they use models of Milky Way-like galaxies and they find inflows $\sim 1-2$~M$_{\sun}$~yr$^{-1}$.
The latter have lower absolute values than in our case, owing to the different conditions of the gas, i.e. different surface density, surface star formation rates, velocity dispersion, etc., but the main difference remains the steady inflows found by \citet{goldbaum+15} in gravito-turbulent disc models.
When we average $\dot{M}(R)$ over time between $z=8.1$ and $z=6.5$, we find large positive and negative fluctuations around zero, hinting against a net and continuous mass inflow due to gravito-turbulence, consistently with the previous arguments.
However, we note that we do not have enough time resolution in the dumped snapshots to firmly assess the convergence of the timely-averaged $\dot{M}(R)$ around zero across the entire disc.

Close to the centre, the gaseous disc becomes proportionally thicker, with an increasing aspect ratio $H / R \sim \sigma_{\rm g} / V_{\phi} \gtrsim 1$ at radii $R \lesssim 300-500$~pc, as shown in the lower panel of Figure \ref{fig_mdot_profile}.
Therefore, the impact of three-dimensional turbulence in the mass flow close to the centre may also affect the angular distribution of moving matter and the direction of streaming gas that might eventually reach the galactic nucleus.
We show that in Figure \ref{fig_mdot_map} through Mollweide angular projections of the mass flow rate per unit of steradian, ${\rm d}\dot{M} / {\rm d}\Omega$, at different redshifts and at different radial distances from the centre.
Specifically, we select gas particles within spherical test shells with radius $r$ and thickness $\Delta r$.
Then, we tessellate the sphere by means of the {\sc healpix} algorithm\footnote{For further information, see \url{http://healpix.sourceforge.net/}. We use the {\sc python} implementation {\sc healpy}, freely available at \url{https://github.com/healpy/healpy}.} and we calculate the mass flow through each angular tassel as ${\rm d}\dot{M} / {\rm d}\Omega = \Delta\Omega^{-1} \Delta r^{-1} \sum_j m_j v_{j, r}$, where $\Delta \Omega = 4 \pi / N$ is the solid angle of the $N$ equal tassels, and the sum extends to the particles within each tassel.

The (instantaneous) mass flow is largely anisotropic on several scales.
Most of the mass flow occurs through the disc plane, including both inflows and outflows at $\gtrsim 5$~M$_{\sun}$~yr$^{-1}$.
On large scales $\sim 1$~kpc (the same order as the disc size), significant inflows and outflows proceed through ``pockets'' of gas localised in solid angle from medium ($\sim 30\degr$) to high ($\gtrsim 60\degr$) latitudes both above and below the disc plane.
This is consistent with the large scale behaviour of the gas triggered by the interplay of stellar feedback and gravity discussed in Section \ref{sec_disc}; indeed, most of the inflowing gas at latitudes far from the disc plane typically has low mass-weighted mean temperature $\lesssim 10^4$~K, while the contrary is true for the outflowing gas, likely pushed away by supernova blast waves.
At intermediate scales $\sim 500$~pc, the inflow-outflow episodes are mostly confined around the disc plane up to latitudes as high as $\sim 30-40\degr$, consistently with the typical thickness of the gaseous disc $\sim 100-300$~pc as seen on the scale of the test shell, while fewer streams of gas cross the test shell at higher latitudes.
On even smaller scales $\sim 100$~pc, the test shell is almost entirely embedded in the thick gaseous disc.
At this scale, the dynamics of the gas within the galactic disc is dominated by turbulent motion and not by rotation, as hinted by the lower panel of Figure \ref{fig_mdot_profile}.
As therefore expected, the mass flow on small scales is highly anisotropic, with well-defined inflow-outflow regions representing cross sections through the test shell of moving over-dense gas clouds at every latitude.
Consistently with Figure \ref{fig_mdot_profile}, the integrated ${\rm d} \dot{M} / {\rm d} \Omega$'s shown in Figure \ref{fig_mdot_map} effectively decrease from larger to smaller scales; however, the value of the net mass flow is different because the former is integrated over a cylindrical shell selecting the flow along the disc plane, while the latter represents the radial flow through a spherical shell.


\section{Discussion and conclusions} \label{sec_4}

In this paper we present the results from PonosHydro, a high-resolution, zoom-in cosmological simulation meant to model the early evolution of a present-day massive galaxy down to $z \sim 6$, whose global properties appear to be consistent with the available data for galaxies of similar stellar masses at those redshifts \citep{iye+06,bradley+12,watson+15}.
Specifically, we study the assembly of the galaxy during its first starburst phase, before the supermassive black hole can exert significant feedback and quench star formation.
We focus on the properties of the interstellar medium and the transport of mass across the galactic disc and study the conditions that determine the early evolution of the central regions of present day massive galaxies.

\begin{figure}
\begin{center}
\includegraphics[width=\columnwidth]{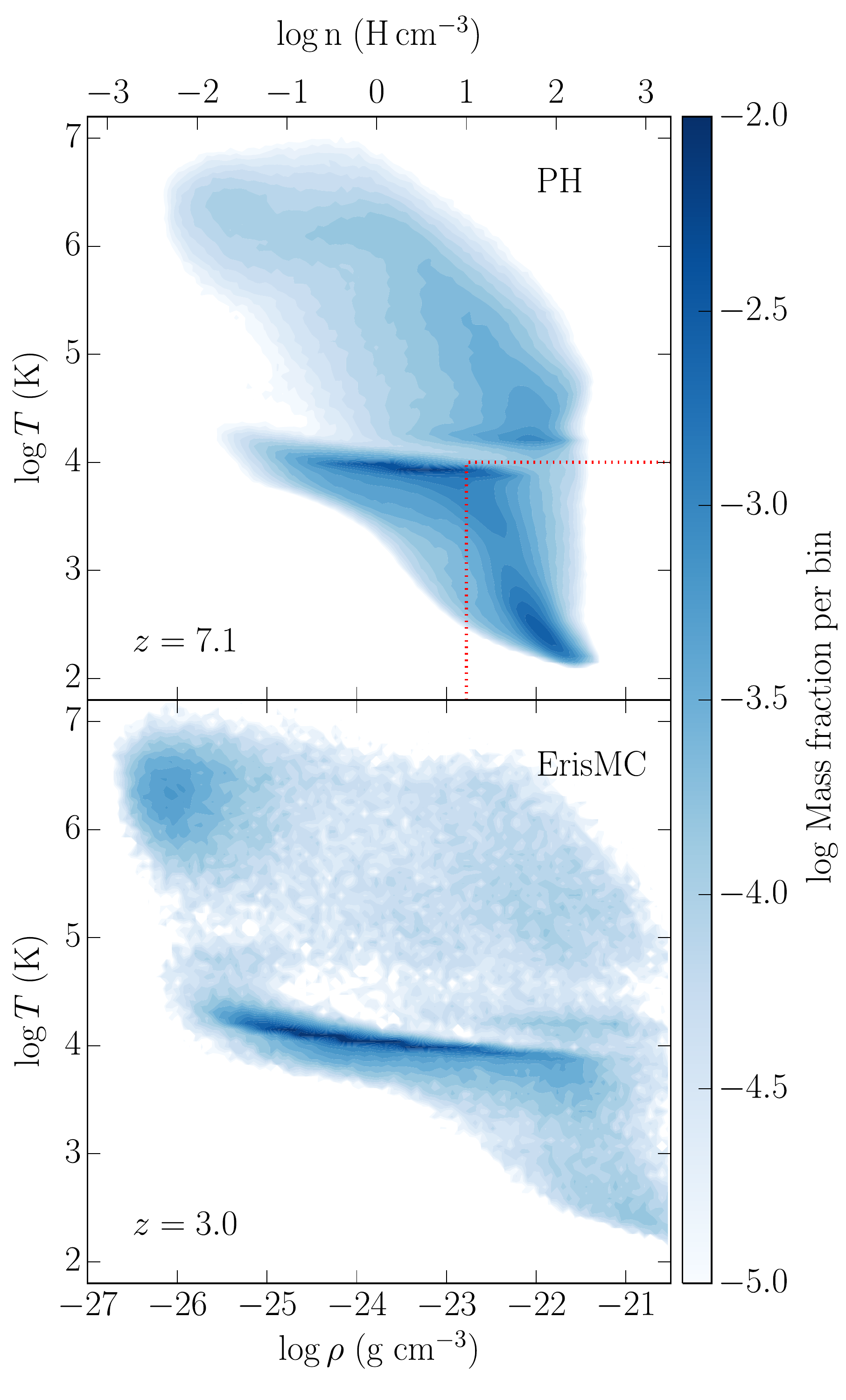}
\caption{Comparison of the phase diagram of run PH at $z=7.1$ (upper panel) and ErisMC at $z=3$ (lower panel).
The colour bar shows the logarithm of the mass fraction per bin in the density-temperature plane.
The red dotted line in the upper panel marks the density and star formation thresholds adopted in run PH.
}
\label{fig_phase_diagram}
\end{center}
\end{figure}

Before we discuss the potential implications of our findings, we briefly comment on the possible shortcomings of our calculations.
Our results may be subject to the feedback model that we have used and this could in principle quantitatively affect our conclusions, at least to some extent.
Various feedback schemes differently change the temperature of the gas locally after supernovae explosions.
The delayed-cooling blast wave feedback produces gas at $\sim 10^5$~K and at densities $\sim 10-100$~H~cm$^{-3}$ by injecting energy in the surrounding of the star-forming regions and preventing them from cooling.
Before this gas expands adiabatically and gets blown away, it significantly contributes to the global stability of the galaxy against the onset of gravito-turbulence and fragmentation.
While the contribution of this phase might be artificially enhanced by the feedback scheme, we argue that the physical conditions of run PH (in particular, the high specific star formation rate and the relatively small mass) are more prone to form a warm and dense gas phase than in local disc galaxies.
Figure \ref{fig_phase_diagram} shows the density-temperature diagram of the gas within 3 and 10 physical kpc around the main galaxy in run PH at $z=7.1$ and ErisMC\footnote{ErisMC is a simulation of a Milky Way-like galaxy that adopts a sub-grid model very similar to ours and therefore it allows to compare the thermodynamics of the gas in different physical conditions, though with the same feedback scheme. For further details about ErisMC, see \citet{shen+12}.
Unfortunately, a snapshot at redshift lower than 3, which might have strengthened our considerations providing a more quiescent galaxy, is not available.} at $z=3$, respectively.
Despite that both simulations show some amount of gas around $10-100$~H~cm$^{-3}$ and $\sim 10^5$~K, this accounts for an order of magnitude less mass fraction in the more quiescent ErisMC, that has almost ten times smaller specific star formation rate \citep{shen+12}.
Similar results have been obtained with different feedback models (see e.g. Fig. 11 of \citealt{hopkins+12b}, where the warm phase is more sub-dominant compared to the cold phase at $\sim 100$~H~cm$^{-3}$ in a Milky Way-like galaxy than in a high-$z$-like galaxy).
Moreover, the latest generation of stronger feedback models, designed to capture well the run between stellar mass and halo mass across cosmic scales and epochs, tend to produce lower density, thicker discs by driving more powerful and hot ($T \gtrsim 10^7$~K) gas outflow \citep{hopkins+14,keller+14}. 
Such galactic discs would likely be less gravitationally unstable and yet turbulent in the gas component (e.g. \citealt{hopkins+12b,hopkins+12,mayer+16}).
Therefore, we conclude that a different and stronger feedback model would likely lead to qualitatively similar conclusions, namely that the gaseous discs of typical star-forming $z \sim 6$ galaxies would be maintained turbulent and stable against gravitational fragmentation by feedback, albeit future tests with different feedback schemes could better assess potential differences.

Two main processes have been often advocated in the literature to shape the global dynamics of the interstellar medium: gravitational instability and supernova feedback.
In Section \ref{sec_3} we discussed the features of the gaseous disc of run PH in terms of star formation, outflows, turbulence, mass transport, and gravitational stability, arguing that feedback likely plays a dominant role in the early evolution of a typical $z \sim 6-7$ galaxy.
This seams to be somewhat different from what is usually expected both at low ($z \sim 0$) and intermediate ($z \sim 2$) redshift.
Recently, \citet{goldbaum+15,goldbaum+16} have thoroughly explored the relative role of gravity and stellar feedback with controlled simulations of present day Milky Way-like galaxies.
Consistently with previous results (e.g. \citealt{agertz+09,bournaud+10,agertz+15}), they find that stellar feedback is important to locally regulate star formation and to create a multi-phase interstellar medium, but the galaxy nonetheless settles to a gravito-turbulent state with a Toomre parameter $Q \sim 1$ that mostly controls the velocity dispersion and the mass transport through the disc.

Those results match the observations of nearby spiral galaxies with low star formation rates (e.g. \citealt{tamburro+09,bagetakos+11}).
They are conceptually similar to what has been often argued for massive $\sim 10^{11}$~M$_{\sun}$, gas-rich galaxies at $z \approx 2$, i.e. star forming galaxies that have not been quenched yet and are likely the progenitors of the most massive quiescent galaxies at $z = 0$, undergoing so called violent disc instability, i.e. gravitational instability that leads to the formation of massive star-forming clumps (e.g. \citealt{dekel+09,ceverino+10,mandelker+14,inoue+16}; but see also \citealt{hopkins+12,tamburello+15}).
In this respect, massive discs at $z \approx 2$ would be the most extreme manifestation of the 
``cold and turbulent'' regime that eventually leads to gravitational instability and fragmentation in the gas, since they are already as massive as the most massive discs in the local Universe but proportionally more gas rich, perhaps because they appear near the peak of the cosmic star formation history \citep{madau+14}.
Motivated by this analogy, \citet{goldbaum+16} have proposed that gravitational instability is the dominant process setting mass transport and fuelling star formation over cosmic time.
However, our results suggest that this might not be the case for typical $z \sim 6-7$ galaxies with stellar and gas mass $\gtrsim 10^{9}$~M$_{\sun}$, where the disc dynamics and mass transport seems to be significantly influenced by stellar feedback.
Indeed, even in the favourable conditions of massive gaseous discs at $z \approx 2$, a phase of violent disc instability with its associated gravito-turbulent can or cannot occur depending on how effective is the feedback model at heating the gas and generating mass-loaded outflows.
Recent work has shown that modern strong feedback models tend to suppress gravitational instability and fragmentation, and at the same time that blast wave feedback cannot suppress disc instability when conditions are favourable for its emergence \citep{mayer+16}, corroborating at least the qualitative distinction between the two cases.
Interestingly, \citet{ceverino+16} recently reach analogous conclusions on the role of stellar feedback by looking at a galaxy with stellar mass $M_{\star} \sim 10^9$~M$_{\sun}$ comparable to ours but at $z \sim 1$.
They find a qualitatively similar evolution over time of the Toomre parameter and the $V_{\phi}/\sigma_{\rm g}$ ratio despite they use a different approach to model stellar feedback, i.e. through non-thermal radiation pressure in addition to thermal dump without shut-off cooling.

We thus argue that the dominant role of stellar feedback in the early evolutionary phase of massive galaxy progenitors is likely controlled by the combination of the high specific star formation rate ($\gtrsim 5$~Gyr$^{-1}$) and of the relatively low mass at $z > 5$ ($\sim 10^{9}$~M$_{\sun}$).
The first favours the impact of stellar feedback on the interstellar medium, while the latter proportionally weakens the dynamical role of gravity to lead to instabilities and eventually fragmentation.
Those specific star formation rates are expected for galaxies on the main star forming sequence at $z > 5$ (Figure \ref{fig_SFH}; \citealt{schreiber+15,tasca+15}); as our galaxy is consistent with the main sequence, this suggests that the ``hot and turbulent'' regime that we characterise here could be typical of star forming galaxies at $z > 5$ with baryonic/stellar  masses comparable to ours.
These should be fairly typical galaxies, as recent surveys begin to find \citep{bradley+12,capak+15,watson+15}.
In particular, recent ALMA observations by \citet{maiolino+15} tend to qualitatively support the idea that stellar feedback has a dominant role in the early assembly of normal star-forming galaxies at $z\sim 6- 7$.
This has immediate observational implications, as we would predict a significant amount of warm/hot gas
with temperature $5\times 10^{4} \lesssim T/{\rm K} \lesssim 5 \times 10^{5}$ inside and around the disc, possibly $\sim 0.1$ of the gas mass.
Note that these are temperatures more akin to the circumgalactic medium distributed in the virial volume around galaxies at low redshift \citep{werk+13}, but in our case it would be inside or surrounding the galactic disc.

Possible analogues in the local Universe may serve as preliminary test bed for our predictions.
Those might be low-mass starburst galaxies, such as the prototypical M82, that has stellar mass and star formation rate rather similar to the main galaxy of run PH (with a factor 3-4 lower specific star formation rate due to the higher stellar mass; e.g. \citealt{forster+03,greco+12}).
While gas densities are expected to be lower at $z=0$, M82 might also host a significant fraction of warm/hot, turbulent gas in its disc, at least in the central kpc where the starburst is ongoing.
This seems to be confirmed by observations (e.g. \citealt{griffiths+00}) and at least in qualitative agreement with our results since gas temperature and phases found in our simulations would be somewhat dependent on the specific feedback model.
Detailed characterisation of the warm/hot interstellar medium in low mass starburst galaxies could thus provide useful constraints to test our scenario.

In this ``hot and turbulent'' regime, the mass transport is influenced by intense and clustered stellar feedback episodes.
As a result, the gas flow through the disc is fluctuating and anisotropic, with no sustained coherent gas inflow within the disc.
A coherent circumnuclear disc, which can be a way to funnel accretion towards the ultimate
stage of the accretion disc, is not clearly seen to form at $\sim 100$~pc scales (though this would be barely resolved at our resolution).
This might have some implications for the feeding of a central massive black hole \citep{gabor+13,dubois+15}.
On one hand, mean inflow rates could be small; however, episodic accretion events at high rates could occur through the infall of massive gas clouds, as we observe inflow rates that can occasionally peak at $\gtrsim 5$~M$_{\sun}$~yr$^{-1}$.
Nonetheless, if super-Eddington accretion is assumed, recent models show that even episodic accretion is enough for the rapid growth of central black holes (e.g. \citealt{lupi+16,pezzulli+16}).
Accretion may also occur in an anisotropic way, with large fluctuations in the angular momentum of the accreting matter, which would have implications for the nature of the accretion disc itself, if any, and for the evolution of the spin of the central black hole.
However, we defer additional speculations on the evolution of a massive black hole in such environments to a forthcoming investigation.

As steady central gas inflows are not sustained, bulge/spheroid formation from dynamical and/or secular
disc instabilities are unlikely to take place (e.g. \citealt{guedes+13}).
Indeed, the disc of PonosHydro remains nearly bulgeless for the whole simulation (see Figure \ref{fig_rot_curve}).
However, we know from lower resolution runs going to $z=0$ that the galaxy will develop a dominant spheroid at lower redshifts as it grows to become a massive early-type galaxy (Fiacconi et al., in preparation).
Since it has several mergers occurring at later times ($2 < z< 4 $; \citealt{fiacconi+16}), it is likely that such mergers will be the dominant driver of spheroid growth \citep{fiacconi+15}.
However, if the ``hot and turbulent'' regimes characterises main sequence galaxies at $z > 5$, this would imply that bulge formation may occur after the first billion year of evolution, possibly post-dating the growth of the massive black hole at the centre (see also \citealt{dubois+15} and \citealt{habouzit+16}).
Therefore, we predict that gas-rich star forming discs at $z > 5$ should not host a significant bulge.

The exploration described in this paper leads to interesting predictions about the early assembly of massive galaxies.
However, our interpretations remain rather speculative, both from the theoretical and the observational point of view.
On one hand, future simulations including different subgrid models are necessary to quantitatively assess the different nature of galaxies at low and high redshifts, also comparing the results from codes with intrinsically different treatments of hydrodynamics (e.g. \citealt{kim+16}).
On the other hand, more detailed characterisation of high redshift galaxies are starting to be available from current observational facilities (e.g. ALMA).
However, it is going to be in the next future that forthcoming observatories (e.g. JWST, E-ELT) will provide deep enough data to definitely test our predictions and at the same time to better guide the theoretical study of galaxies at the cosmic dawn.


\section*{Acknowledgements}

We thank the anonymous Referee for constructive comments that helped us to improve the quality of the paper.
We acknowledge useful discussions with Arif Babul, Rychard Bouwens, Nick Gnedin, Raffaella Schneider, Sijing Shen, and Debora Sijacki.
We thank Oliver Hahn and the AGORA collaboration for help with the initial conditions of the simulations.
The simulations have been run on the ZBOX4 cluster at the University of Zurich and on the Pitz Dora cluster at CSCS, Lugano.
We acknowledge the use of the {\sc python} package {\sc pynbody} (\citealt{pontzen+13}; publicly available at \url{https://github.com/pynbody/pynbody}) in our analysis for this paper.
D.F. is supported by the Swiss National Science Foundation under grant \#No. 200021\_140645.
D.F. also acknowledges support by ERC Starting Grant 638707 ``Black holes and their host galaxies: coevolution across cosmic time''.
Support for this work was provided to P.M. by the NSF through grant AST-1229745, and by NASA through grant NNX12AF87G.
P.M. also acknowledges a NASA contract supporting the WFIRST-EXPO Science Investigation Team (15-WFIRST15-0004), administered by GSFC, and thanks the Pr\'{e}fecture of the Ile-de-France Region for the award of a Blaise Pascal International Research Chair, managed by the Fondation de l'Ecole Normale Sup\'{e}rieure.


\bibliographystyle{mnras}
\bibliography{ponos_high_z}


\appendix

\section{Resolution tests} \label{appendix_resolution_tests}

\begin{figure*}
\begin{center}
\includegraphics[width=16cm]{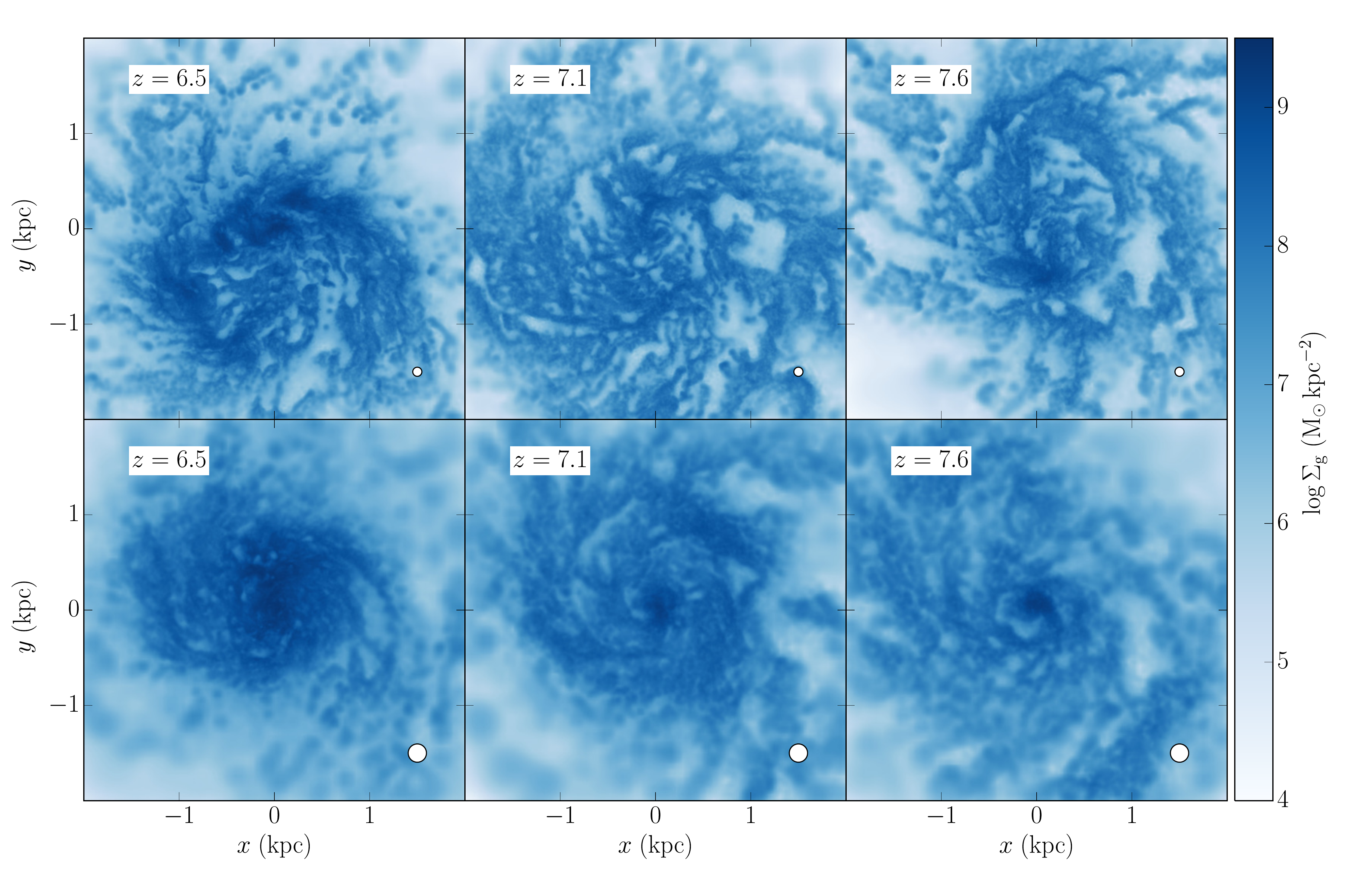}
\caption{Comparison of the surface density of the gas within the galactic disc between runs PH (upper row) and PH\_LR (bottom row).
From left to right, we show the data at $z=6.5$, 7.1, and 7.6, respectively.
The white circle at the bottom-right corner of each panel shows the size of the softening in physical units.
}
\label{fig_res_map}
\end{center}
\end{figure*}

\begin{figure}
\begin{center}
\includegraphics[width=8cm]{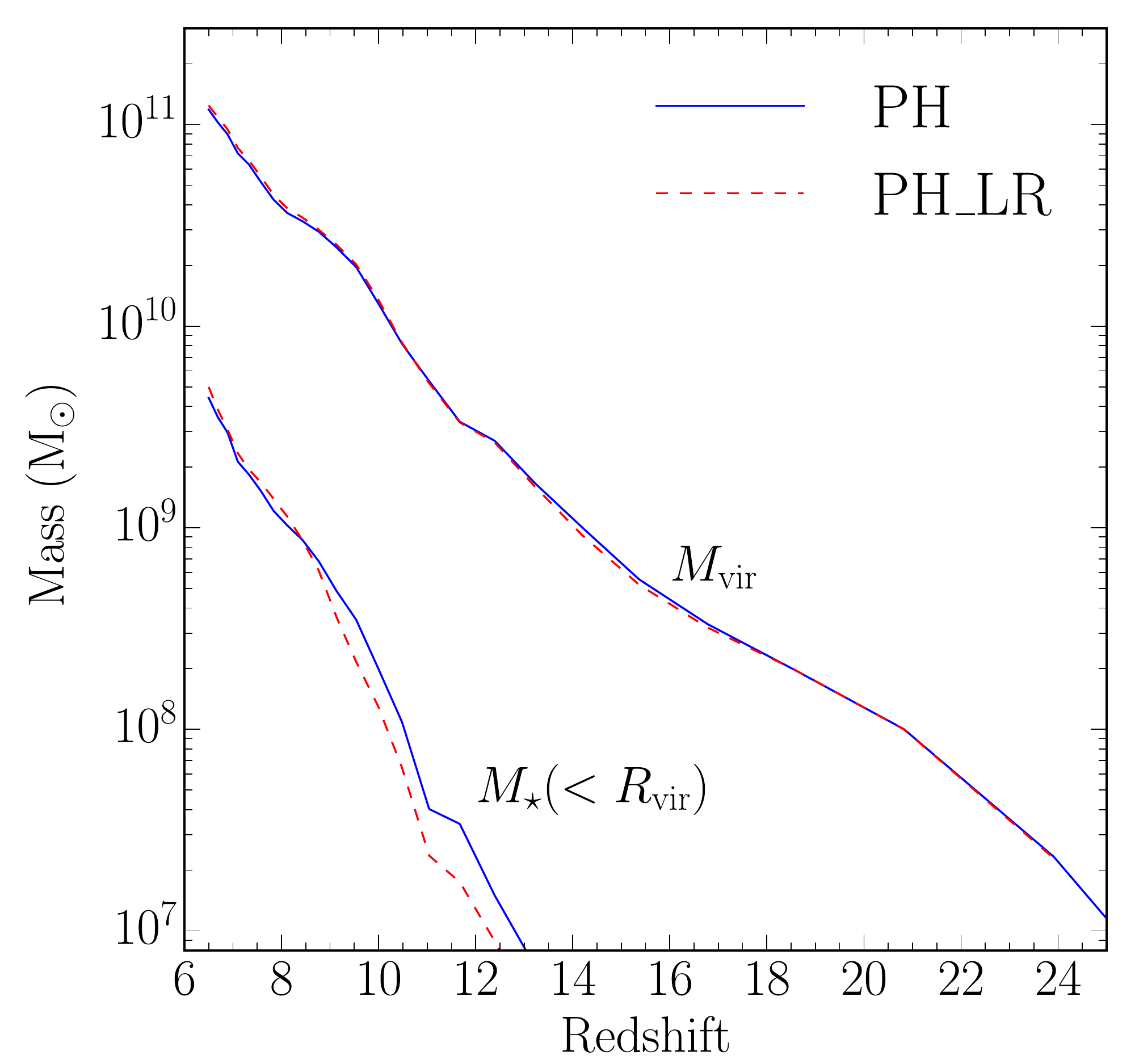}
\caption{Comparison of the redshift evolution of the virial mass $M_{\rm vir}$ and of the total stellar mass $M_{\star}$ within $R_{\rm vir}$ (i.e. main galaxy and subhaloes) for runs PH (blue solid line) and PH\_LR (red dashed line).
}
\label{fig_res_mass}
\end{center}
\end{figure}

We run a twin simulation at lower resolution dubbed ``PH\_LR'', as described in Section \ref{sec_sim_code}.
Figure \ref{fig_res_map} shows a comparison of the surface densities of the galactic disc in runs PH and PH\_LR at $z=7.6$, 7.1, and 6.5.
The two runs produce galaxies that look qualitatively similar, at least between $z=7.6$ and $z=7.1$.
However, we systematically observe that the gas in run PH\_LR is slightly more compact and concentrated at the centre, in qualitative agreement with previous results from controlled simulations to study the angular momentum transport in cooling haloes as a function of resolution \citep{kaufmann+07}.
Moreover, the interstellar medium shows overall smoother features than in PH, as expected given the lower resolution.
The major differences are at $z=6.5$, when the two galaxies also react slightly differently to the perturbation induced by a nearby companion that is going to merge with the main galaxy.

\begin{figure}
\begin{center}
\includegraphics[width=8cm]{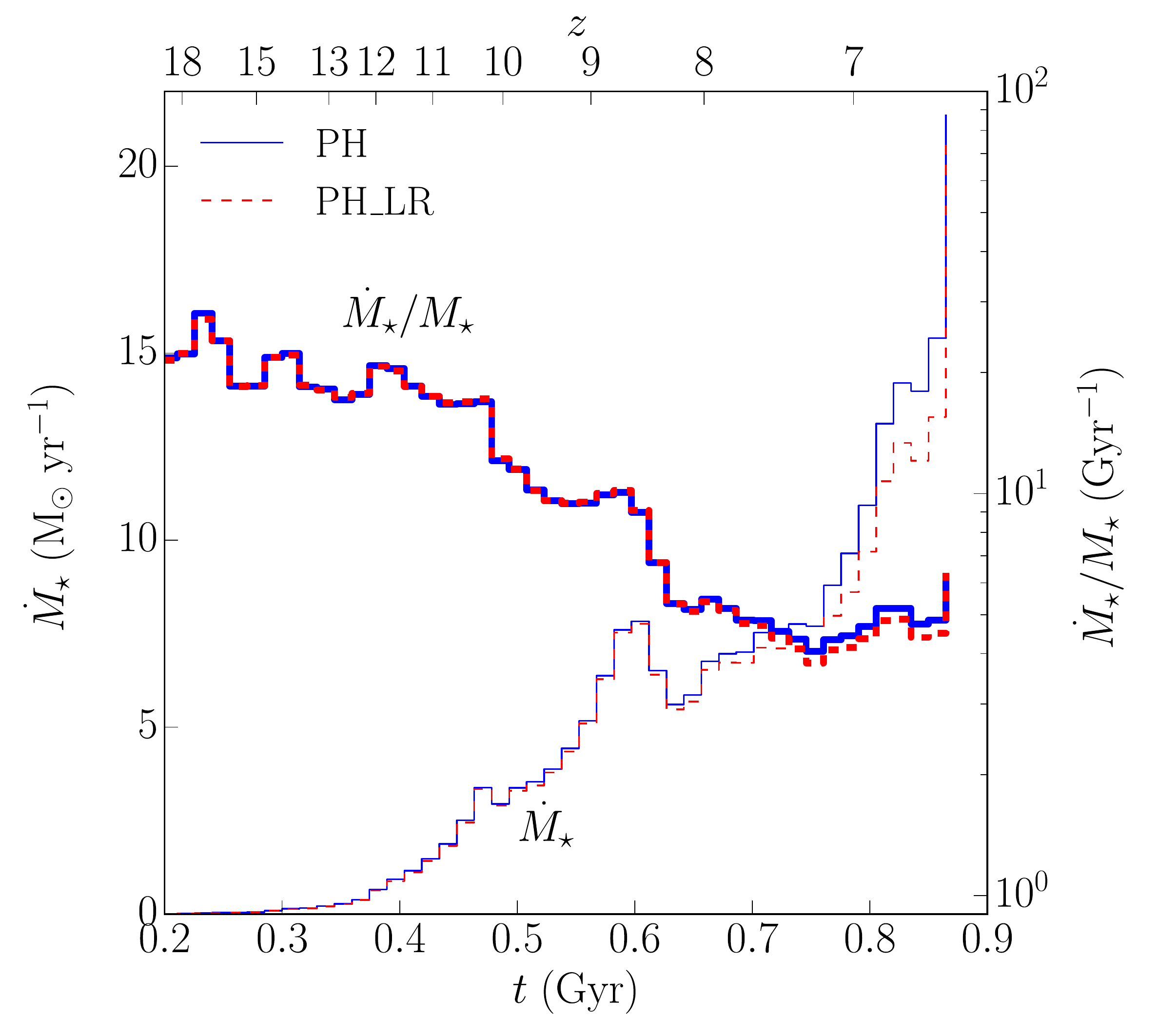}
\caption{Comparison of the evolution of the star formation rate (thin lines, left $y$-axes) and of the specific star formation rate (thick lines, right $y$-axes) between runs PH (blue solid line) and PH\_LR (red dashed line).
}
\label{fig_res_SF}
\end{center}
\end{figure}

More quantitatively, Figure \ref{fig_res_mass} shows the evolution with redshift of both the virial mass, $M_{\rm vir}$, and the total stellar mass (main galaxy and satellites) within the virial radius, $M_{\star}(<R_{\rm vir})$.
The growth of $M_{\rm vir}$ with time is almost indistinguishable between the two runs, while the stellar mass shows mild differences at $z>9$, with $M_{\star}(<R_{\rm vir})$ in run PH\_LR being roughly a factor $\lesssim 2$ smaller than in run PH, but this is also possibly related to the evolution of the satellites.
Similarly, Figure \ref{fig_res_SF} shows the comparison of the (specific) star formation rate as a function of time for the main galaxy.
The two runs show little differences at the two resolutions, suggesting that the combination of the subgrid models with the evolution of the galaxies might have reached reasonable convergence.

\begin{figure}
\begin{center}
\includegraphics[width=8cm]{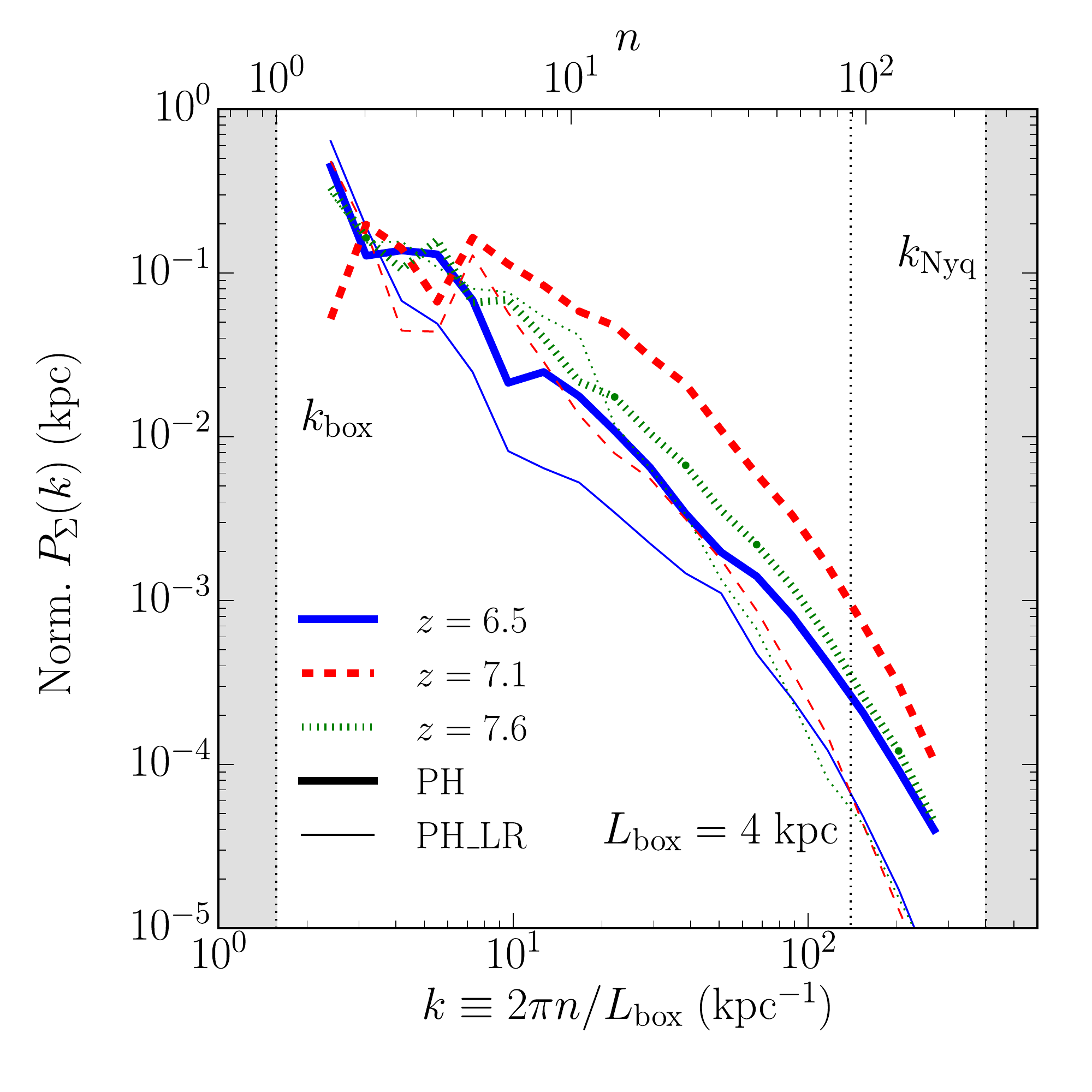}
\caption{Comparison of the normalised surface density power spectra for runs PH (thick lines) and PH\_LR (thin lines).
Blue solid, red dashed, and green dotted lines refer to $z = 6.5$, 7.1, and 7.6, respectively.
The left and right grey shaded regions mark the scale $k$ associated to the box size ($L_{\rm box} = 4$~kpc) and to the Nyquist frequency associated to the grid spacing (see Section \ref{sec_turbulence} for additional details).
}
\label{fig_res_SD_PS}
\end{center}
\end{figure}

Finally, Figure \ref{fig_res_SD_PS} compares the surface density power spectra of run PH\_LR with the fiducial one (see Section \ref{sec_turbulence} for further details).
The power spectra of the low resolution case also shows a similar behaviour with two nearly power law for $k$ smaller and larger than $\sim 20-30$~kpc$^{-1}$; however, the exponents tend to be systematically slightly lower (i.e. steeper power law) for PH\_LR.
This difference is more significant for large $k$ modes, where the low resolution run shows less power than the fiducial run.
This is likely related to the lower mass and spatial resolution that tends to suppress over-densities at small scales.
On the other hand, the largest scales ($k < 10$~kpc$^{-1}$) seem to reach reasonable convergence, at least at $z = 7.1$ and 7.6, while at $z=6.5$ the difference is larger, possibly because of the different response of the main galaxy in the two runs to the perturbing satellite visible in Figure \ref{fig_merger_tree}.
Despite these differences, the qualitative agreement in the shape of $P_{\Sigma}$ for the high- and low-resolution simulations suggests that the interstellar medium reaches a similar configuration in the two cases.
In addition, also the typical value of the gas velocity dispersion $\sigma_{\rm gas} \sim 40$~km~s$^{-1}$ is comparable among the two runs.
However, due to the lower resolution (i.e. the lower number of gas particles in the galactic disc), we cannot compare the velocity power spectra in 512~pc cubic boxes directly.


\label{lastpage}

\end{document}